\documentclass[iop]{emulateapj}
\usepackage{color}
\usepackage{rotating}
\usepackage{soul}

\begin{document}
  
\title{ Correcting Type Ia Supernova Distances for Selection Biases and \\ Contamination 
 in Photometrically Identified Samples }
   
\author{
R.~Kessler\altaffilmark{1,2},
D.~Scolnic\altaffilmark{1,3}
}
 
\altaffiltext{1}{Kavli Institute for Cosmological Physics, University of Chicago, Chicago, IL 60637, USA}
\altaffiltext{2}{Department of Astronomy and Astrophysics, University of Chicago, 5640 South Ellis Avenue, Chicago, IL 60637, USA}
\altaffiltext{3}{Hubble, KICP Fellow}

% acronyms
% \newcommand{\acro}{{\tt BCD}}
% \newcommand{\acroDef}{Bias Corrected Distances}

\newcommand{\acro}{{\tt BBC}}
\newcommand{\acroDef}{BEAMS with Bias Corrections}
\newcommand{\BEAMS}{{\tt BEAMS}}

\newcommand{\SNANA}{{\tt SNANA}}
\newcommand{\Smu}{{\tt SALT2mu}}
\newcommand{\SALTII}{{\sc SALT-II}}
\newcommand{\SDSS}{SDSS-II}
\newcommand{\PS}{Pan-STARRS1}
\newcommand{\Diff}{{\tt DiffImg}}
\newcommand{\hotPants}{{\tt hotPants}}
\newcommand{\autoScan}{{\tt autoScan}}
\newcommand{\DESSN}{DES-SN}
\newcommand{\wCDM}{$w$CDM}
\newcommand{\lowz}{low-$z$}

% plain English
\newcommand{\DES}{Dark Energy Survey}
\newcommand{\Spec}{Spectroscopic}
\newcommand{\spec}{spectroscopic}
\newcommand{\specy}{spectroscopically}
\newcommand{\obs}{observation}
\newcommand{\obss}{observations}
\newcommand{\eff}{efficiency}
\newcommand{\ineff}{inefficiency}
\newcommand{\effs}{efficiencies}
\newcommand{\unc}{uncertainty}
\newcommand{\uncs}{uncertainties}
\newcommand{\GN}{Gaussian normalization}
\newcommand{\lh}{likelihood}
\newcommand{\lhs}{likelihoods}
\newcommand{\post}{posterior}  
\newcommand{\Prob}{Probability}
\newcommand{\prob}{probability}
\newcommand{\probs}{probabilities}
\newcommand{\biasCor}{BiasCor}
\newcommand{\con}{contamination}
\newcommand{\Con}{Contamination}

% math symbols
\newcommand{\OL}{\Omega_{\Lambda}}
\newcommand{\OM}{\Omega_{\rm M}}
\newcommand{\Pfit}{P_{\rm fit}}
\newcommand{\Mo}{M_0}
\newcommand{\DMz}{\Delta M_0}

% asymmetric Gaussians
\newcommand{\cbar}{\bar{c}}
\newcommand{\xbar}{\bar{x}_1}
\newcommand{\sigminus}{\sigma_{-}}
\newcommand{\sigplus}{\sigma_{+}}

\newcommand{\wref}{w_{\rm ref}}
\newcommand{\wfit}{w_{\rm fit}}
\newcommand{\wdata}{w_{\rm data}}
\newcommand{\wrefVal}{-1}

\newcommand{\OMref}{\Omega_{\rm M,ref}}
\newcommand{\OMfit}{\Omega_{\rm M,fit}}
\newcommand{\OMrefVal}{0.3}
\newcommand{\mutrue}{\mu_{\rm true}}
\newcommand{\mumodel}{\mu_{\rm model}}
\newcommand{\mumodeli}{\mu_{{\rm model},i}}
\newcommand{\mumodelz}{\mu_{{\rm model},z}}
\newcommand{\murefz}{\mu_{{\rm ref},z}}

\newcommand{\prior}{P(\theta)} 
\newcommand{\chisqHD}{\chi^2_{\rm HD}}
\newcommand{\chisqDelta}{\chi^2_{\Delta}}
\newcommand{\chisqred}{\chi^2_{\rm red}}
\newcommand{\chisqdof}{\chi^2/{\rm dof}}

\newcommand{\sigmu}{\sigma_{\mu}}
\newcommand{\sigmui}{\sigma_{\mu,i}}
\newcommand{\sigmuzerr}{\sigma_{\mu}^z}  % sigma_mu from redshift error
\newcommand{\sigmuz}{\sigma_{\Delta\mu,z}}       % error on \Delta_mu
\newcommand{\Rsig}{R_{\sigma}}
\newcommand{\sigmuData}{\sigmu^{\rm data}}
\newcommand{\sigmuSim}{\sigmu^{\rm sim}}
\newcommand{\sigvpec}{\sigma_{\rm vpec}}
\newcommand{\sigint}{\sigma_{\rm int}}
\newcommand{\sigcoh}{\sigma_{\rm coh}}
\newcommand{\sigw}{\sigma_w}
\newcommand{\Cint}{\Sigma}
\newcommand{\Trest}{T_{\rm rest}}
\newcommand{\aSIM}{\alpha_{\rm SIM}}
\newcommand{\bSIM}{\beta_{\rm SIM}}	

\newcommand{\Ndof}{N_{\rm dof}}
\newcommand{\NdofIa}{N_{\rm dof}^{\rm Ia}}
\newcommand{\DMU}{\Delta_{\mu}}
\newcommand{\DMUi}{\Delta_{{\mu},i}}
\newcommand{\DMUz}{\Delta_{{\mu},z}}
\newcommand{\vecD}{\vec{D}_{\mu,z}}

\newcommand{\Ndata}{N_{\rm data}}
\newcommand{\Nz}{N_z}

\newcommand{\Dz}{{\rm D}_z}
\newcommand{\Dx}{{\rm D}_{x1}}
\newcommand{\Dc}{{\rm D}_c}
\newcommand{\PIa}{P_{\rm NN,Ia}}
\newcommand{\Ptot}{P_{\rm tot}}
\newcommand{\Pratio}{ P_{\rm CC/Ia}}
\newcommand{\SCC}{S_{\rm CC}}
\newcommand{\EffIa}{{\rm Eff}_{\rm Ia}}
\newcommand{\Nfitpar}{N_{\rm fitPar}}
\newcommand{\COV}{{\cal C}}

\newcommand{\mucor}{\mu^{\star}}
\newcommand{\mBcor}{m_B^{\star}}
\newcommand{\xcor}{x_1^{\star}}
\newcommand{\ccor}{c^{\star}}

\newcommand{\LH}{\mathcal L}
\newcommand{\DIa}{D_{\rm Ia}}
\newcommand{\DCC}{D_{\rm CC}}
\newcommand{\dmB}{\bar\delta_{m_B}}
\newcommand{\dx}{\bar\delta_{x_1}} 
\newcommand{\dc}{\bar\delta_{c}}
\newcommand{\dmu}{\bar\delta_{\mu}}

\newcommand{\Wi}{W_{{\rm Ia},i}}
\newcommand{\zbin}{\bar{z}_{\rm bin}} 
\newcommand{\RCC}{R_{\rm CC}}

\newcommand{\COSVEC}{\vec{C}}

% - - - - - - - - - - - - - - 
% numbers/results
\newcommand{\zmax}{1.10}        % max redshift for SN samples
\newcommand{\NSAMPLE}{15}
\newcommand{\NSIMDATA}{10{,}000}
\newcommand{\NSIMTRAIN}{70{,}000}
\newcommand{\NSIMBIASCOR}{500{,}000}
\newcommand{\NSIMDATASUM}{900{,}000}  % 6 test, each with 15 samples of 10k

% biasCor info
\newcommand{\CELLSIZEz}{0.05}
\newcommand{\CELLSIZEx}{0.50}
\newcommand{\CELLSIZEc}{0.05}
\newcommand{\NBINz}{22}   % number of biasCor z-bins (excluding buffer bin at z>1.1)
\newcommand{\NBINx}{16}   % number of biasCor x1 bins   
\newcommand{\NBINc}{16}   % number of biasCor c bins
\newcommand{\NPERCELLMIN}{3}  % min number of biasCor events per cell
\newcommand{\NCELLMIN}{3}        % at least this many cells for interpolation

%\newcommand{\NBTOT}{5888}  % product of above
%\newcommand{\NBUSE}{2000}  % approx number of used biasCor bins

% NN numbers here are for nominal CC rate
\newcommand{\SNIAeffNN}{0.997}           % NN efface for SNIa
\newcommand{\SNIAineffNN}{0.003}        % 1- above
\newcommand{\CCfracBoxCuts}{0.054}    % true CC/tot ratio after box cuts
\newcommand{\CCfracNN}{0.015}             % idem after NN

% add 'x' for x3 CC rate
\newcommand{\SNIAeffNNx}{0.992}           % NN efface for SNIa
\newcommand{\SNIAineffNNx}{0.008}        % 1- above
\newcommand{\CCfracBoxCutsx}{0.146}    % true CC/tot ratio after box cuts
\newcommand{\CCfracNNx}{0.035}             % idem after NN

\newcommand{\amin}{0.12}   % grid alpha used for biasCor
\newcommand{\amax}{0.16}  % idem
\newcommand{\bmin}{2.8}      % idem for beta
\newcommand{\bmax}{3.6}
\newcommand{\agen}{0.14}    % generated alpha
\newcommand{\bgen}{3.2}     %  generated beta
\newcommand{\sigCOH}{0.13}      %  generated sigInt for COH model
\newcommand{\sigGTEN}{0.076}  % sigInt for G10 scatter model
\newcommand{\sigintVal}{0.1}       % sigInt value from previous result

% ABCD fit results with no CC and no MW
\newcommand{\WBIAS}{-0.002\pm 0.004}  % ABCD fit without CC, without MW, wdata=-1
\newcommand{\WCHISQRED}{0.5}           % reduced chi2 for WBIAS
\newcommand{\sigwAdd}{\sigw/7}              % combined Deep+shallow
\newcommand{\sigwAddDeep}{\sigw/12}    % for deep fields
\newcommand{\sigwAddShallow}{\sigw/5}    % for shallow fields

\newcommand{\wBiasAvgCC}{$0.006\pm 0.002$}  % ABCD grand-average w-bias over 3 scatter models and CCx1 and CCx3

\email{kessler@kicp.uchicago.edu}
\submitted{Accepted by ApJ, January 12, 2017}

\begin{abstract}
We present a new technique to create a bin-averaged Hubble diagram (HD) 
from  photometrically identified SN~Ia data.
The resulting HD is corrected for selection  biases and contamination from 
core-collapse (CC) SNe, and can be used to infer cosmological parameters.
This method, called ``{\acroDef}" (\acro), includes two fitting stages. 
The first \acro\ fitting stage 
uses a \post\ distribution that includes multiple SN \lhs,
a Monte Carlo simulation to bias-correct the fitted \SALTII\ parameters,
and  CC probabilities determined from a machine-learning technique.
The \acro\ fit determines
1) a bin-averaged HD (average distance vs. redshift), and
2) the nuisance parameters $\alpha$ and $\beta$,
which multiply the stretch and color  (respectively) to standardize the 
SN brightness. In the second stage,  the bin-averaged HD is
fit to a cosmological model where priors can be imposed.
We perform high-precision tests of the \acro\ method by simulating large 
(150,000 event)  data samples  corresponding  to the 
Dark Energy Survey Supernova Program.
Our tests include three models of intrinsic scatter,  each with two different CC rates.
In the \acro\ fit, the \SALTII\ nuisance parameters $\alpha$ and $\beta$
are recovered to within 1\% of their true values. 
In the cosmology fit, we determine the dark energy equation of state parameter $w$
using a fixed value of  $\OM$ as a prior: averaging over all six tests based on  
$6\times 150,000=\NSIMDATASUM$ SNe, there is a small $w$-bias of \wBiasAvgCC.
Finally, the \acro\ fitting code is publicly available in the \SNANA\ package.

\end{abstract}

% ======================================================
% ======================================================
 \section{Introduction}
 \label{sec:intro}
% ======================================================
% ======================================================
%
The discovery of the accelerating expansion of the universe \citep{Riess98,Saul99} using 
Type Ia supernovae (SN Ia) has  motivated increasingly large transient searches in 
broadband imaging surveys. 
Approximately $1000$ \specy\ confirmed SNe~Ia  
\citep{Conley2011,JLA,Rest2014,Scolnic2014b}
have been combined with measurements of the cosmic microwave background 
\citep{Hinshaw2013,Planck2014} 
to measure the dark energy equation of state parameter ($w$) and today's
matter density ($\OM$).
There are not enough \spec\ resources to dramatically increase the sample size
of {\it confirmed} SNe~Ia,
and therefore several  programs are aiming to acquire very large samples
of photometrically identified SNe~Ia to measure dark energy properties with
increased precision.
These SN programs include the recently completed \PS\  \citep{PS_2002},
the ongoing \DES\ Supernova Program (DES-SN: \citealt{DESSN2012}), and
the Large Synoptic Survey Telescope (LSST:  \citealt{Ivezic2008,LSST_SciBook}),
expected to begin in the next decade.

These photometric samples are expected to include \con\ from core-collapse (CC) SNe, 
and this \con\ must be accounted for in the inference of cosmological parameters. 
Biases from selection effects and  light-curve fitting must also be accounted for.  
In this paper we  
enhance  {\BEAMS}\footnote{{\BEAMS}: ``Bayesian Estimation Applied to Multiple Species'' \citep{Kunz2007}}
and present a new Hubble diagram (HD) fitting method, called ``{\acroDef}" ({\acro}),
for extracting bias-corrected cosmological parameters from a 
photometrically identified SN~Ia sample. 
We test \acro\  on high-quality simulations of the DES-SN program.
The \SALTII\ framework \citep{Guy2010} is used for light curve fitting,
and  a nearest neighbor (NN) method is used for photometric classification.
We assume that an accurate \spec\ redshift is obtained from the host galaxy, 
and ignore the small fraction of wrong SN-host matches described in \cite{Gupta2016}.
\acro\ is not sensitive to the particular method of photometric classification,
but is sensitive to how well the resulting \con\ is modeled by simulations.
Although \acro\ is designed to treat photometrically identified samples,
it is also applicable to \specy\ confirmed samples by simply leaving
out the CC \lh\ term.

An ideal \lh\ approach would fit for all relevant parameters
(cosmology, nuisance, color and stretch population, redshift dependences, etc.)
and at each fitting step where parameters are varied,
a Monte Carlo simulation (MC) would be run to evaluate
the impact from selection effects.
Since the color and stretch \uncs\ are comparable to the width of the
parent distribution, an ideal \lh\ should use parent distributions and
avoid  approximate $\chi^2$ \lhs\ that incorrectly assume
symmetric Gaussian \uncs. 
The practical reality, however, is that with current techniques the
 computing resources  increase 
as one approaches the ideal \lh, and  all fitting implementations so far use 
approximations that speed up the fitting implementation.

Strategies to achieve an  ideal  principled
\lh\ have been developed within the
``Bayesian hierarchical model''  (BHM) framework \citep{March2011,UNITY,Mandel2016,BAHAMAS}.
While  BHM methods rigorously address the \lh\ issues,
practical approximations are made on the bias corrections.
\citet{March2011,Mandel2016} do not include  bias corrections.
\citet{UNITY} do not use simulations and instead describe the selection 
\eff\ with an ad hoc function that includes additional fitted parameters.
\citet{BAHAMAS} use a redshift-dependent distance bias computed
from simulations in \citet{JLA},
but this simulation is not updated as parameters are  varied in the cosmology fit,
nor does it account for biases in the individual \SALTII\ parameters.
In our \acro\ approach we continue with the fundamentally flawed
$\chi^2$ \lh, but incorporate a more accurate simulation technique to correct 
for selection  biases and CC contamination.
\acro\  requires modest computing resources, allowing  rapid iterations
for systematics studies, and it results in small biases
that are an order of magnitude below current \uncs.

Since all SN-cosmology likelihoods include approximations, we do not claim that 
our \acro\ method is superior over BHM methods, but rather complementary in what
is sacrificed for computational \eff.
For any given method, it is important
to accurately measure biases on high-statistics ($> 10^5$) 
samples of realistic simulated SN light curves that are fit with the \SALTII\ model.
We naively expect that smaller biases require more computing time,
and for a particular science analysis there should be enough information
about each method to select the appropriate  compromise. 
For the \acro\ method presented here, we report biases from nearly a 
million simulated supernovae.

Given the approximate nature of the \lh\ approach, 
it is worth mentioning an alternative {\lh}-free approach under development, 
called ``approximate Bayesian computation,'' or ABC
\citep{ABC2013,ABC2016}.  For each set of fitted parameters,
this method relies solely on an accurate simulation to predict 
observations. A key challenge with the ABC method is to define
an optimal metric that quantifies the consistency between
the data and simulation.

Recent SN~Ia cosmology analyses are based on a light curve fit for
each event using the \SALTII\ model, which determines the
best-fit values for the overall amplitude ($x_0$), stretch ($x_1$), and color ($c$).
The stretch and color describe rest-frame properties of the SN~Ia that are
needed to standardize the brightness, and the amplitude describes the 
observed SN brightness and dimming from the distance modulus. 

The ensemble of fitted $\{x_0,x_1,c\}$  values are passed to a second stage
of  HD fitting. The HD fit simultaneously determines the 
cosmological parameters (e.g., $w$ and $\OM$) 
and standardization coefficients, $\alpha$ and $\beta$, which multiply the
stretch and color, respectively, in order to standardize the brightness
and determine a distance for each event.
\citet[hereafter M11]{SALT2mu} introduced  another HD fitting method 
to determine $\alpha$ and $\beta$ without simultaneously fitting for the cosmological parameters, 
and to compute a cosmology-independent distance modulus for each event.
This  ``{\Smu}'' method fixes the cosmological parameters and fits for a 
distance modulus offset  ($\DMUz$) in multiple redshift bins.
The extraction of cosmological parameters can be obtained by fitting
the distance moduli vs. redshift, or fitting $\DMUz$ vs. redshift. 
The advantage of the \Smu\ output is that  a wide variety of cosmology 
models and priors can be employed  in subsequent analyses
without repeatedly fitting for the nuisance parameters ($\alpha,\beta$).

HD chi-squared  fitting based on fitted \SALTII\ parameters is
fundamentally flawed because it does not
account for biases from selection effects,  light curve fitting, and CC \con.
With \specy\ confirmed SN~Ia samples, the effect from biases has been evaluated 
in recent analyses \citep{Conley2011,JLA,Scolnic2014b}  by running the same flawed 
procedure on both data and a simulation, and using the simulated sample to 
measure the average distance modulus bias in redshift bins.
This bias-vs-redshift was applied as a correction to the data in a second-iteration HD fit.

This iterative correction is conceptually flawed for three reasons.
First, \citet[hereafter SK16]{SK2016} have shown that correcting biases as a 
function of redshift is not a full description, and that the proper correction requires 
a 3D function of redshift, stretch, and color.
Second, the standardization parameters $\alpha$ and $\beta$ were
determined  without bias corrections, and then used to simulate
bias corrections vs. redshift.
The third issue is more of an implementation flaw rather than a conceptual flaw: 
simulations in previous analyses were generated with approximate stretch and color 
populations  that were not rigorously determined as in SK16.

The basic concept of \acro\  is to analyze the fitted \SALTII\ parameters ($x_0,x_1,c$)
by maximizing a  \post\ {\prob}  that combines elements from the
1) \BEAMS\ method  from \citet{Kunz2007} and \citet[hereafter H12]{BEAMS},
     to form a \lh\ from the  Ia and CC {\lhs} ,
2) \Smu\ program (M11), to fit for a distance modulus offset in redshift bins, and
3) SNANA simulations \citep{SNANA},  to determine biases.
  The simulation is also used to determine the  shape of the CC \prob\ map as a function 
  of distance modulus and redshift,
  which should be more accurate than the analytical approximation used in H12.

One of the original goals of the \Smu\ program is to convert the fitted \SALTII\
light curve parameters into a distance modulus for each SN~Ia, so that the
resulting HD can be fit with arbitrary cosmology-fitting programs. 
This strategy does not work with CC \con\ because cosmology-fitting programs 
implicitly assume that all events are genuine SNe~Ia.
Using \acro,  however, the fitted distance modulus offset in each 
redshift ($z$) bin is properly corrected for CC \con. 
The resulting $\DMUz$-vs-$z$ function can be fit with arbitrary cosmology models,
maintaining the original spirit of the \Smu\ program. 

As part of developing  \acro, we address a long-standing paradox
about the HD-fit $\chi^2$ formalism. Within the \SALTII\ framework, 
the \unc\ on the distance modulus ($\sigmu$) depends on fitted nuisance parameters 
($\alpha,\beta$) and thus a \GN\ term, ``$-2\log(\sigmu)$,"  
should be added to the $\chi^2$ function that is minimized.
It has been long recognized, however, that  adding the \GN\ term results
in  large biases on the fitted parameters (e.g., see Appendix B of \citet{Conley2011}).
For a \specy\ confirmed SN~Ia sample, we verify that adding the \GN\ term
does indeed result in large biases if bias corrections are ignored. Including
both the \GN\ term along with bias corrections results in good parameter estimates.

The analysis presented here includes only statistical \uncs, in order to check the 
precision of the \acro\ method. We therefore assume that our simulation correctly
predicts CC \con,  measurement noise,  SN~Ia intrinsic scatter, and selection biases.
An analysis of real data, however, must characterize the accuracy
of the simulations and include sources of inaccuracy in the calculation of
systematic \uncs. 

An overview of the paper is as follows. 
We begin with a review of the \SALTII\ framework in \S\ref{sec:SALT2}.
Simulations are described in \S\ref{sec:sim}.
The photometric analysis and NN  method are described in \S\ref{sec:NN}.
The \acro\ method is presented in \S\ref{sec:bcd},
with results in \S\ref{sec:results-I} and \S\ref{sec:results-II}.
A comparison with other HD fitting methods is given in  \S\ref{sec:oldBiasCor},
and we conclude in \S\ref{sec:fin}.
All simulation and analysis codes used in this paper are publicly available in the
\SNANA\ package,\footnote{\tt http://snana.uchicago.edu}
including analysis input files.\footnote{\tt \$SNDATA\_ROOT/sample\_input\_files/KS2016}

% ======================================================
% ======================================================
 \section{Review of \SALTII\  Framework}
 \label{sec:SALT2}
% ======================================================
% ======================================================
%
Here we briefly review  key aspects of the  \SALTII\  model and \Smu\  program, 
which are critical components of \acro.
As described in \S\ref{sec:intro}, the fitted light-curve parameters for each SN  are 
epoch of peak brightness ($t_0$),
amplitude ($x_0$), stretch ($x_1$), and color ($c \simeq B-V$ at $t_0$).
For each event the fitted parameters are used to standardize the SN brightness and
determine a distance modulus using the Tripp relation \citep{Tripp1998},
\begin{equation}
   \mu = m_B + \alpha x_1 - \beta c - \Mo
    \label{eq:Tripp}
\end{equation}
where $m_B = -2.5\log(x_0)$, $\Mo$ is the rest-frame magnitude for an SN~Ia
with $x_1=c=0$, and $\alpha,\beta$ are global nuisance parameters to standardize
the SN~Ia brightness. 

After the light-curve fits, the next step is a global fit comparing each
measured distance modulus ($\mu$) to a model distance that depends on redshift
and cosmological parameters, $\mumodel = -2.5\log(d_L/10{\rm pc})$,
where for a flat \wCDM\ universe ($\OM+\OL=1$),
\begin{eqnarray}
  d_L(z,w,\OM)  & = & (1+z) \frac{c}{H_0}  \int_0^z  \frac{ dz^{\prime}} { E(z^{\prime}) }~,\\
   E(z)  & = & \left[    \OM(1+z)^3  + \OL(1+z)^{3(1+w)}  \right]^{1/2}~.  \nonumber
\end{eqnarray}
The \Smu\ program uses {\tt MINUIT}\footnote{www.cern.ch/minuit}
to minimize the HD chi-squared function
\begin{eqnarray}
   \chisqHD & = & \sum_i ( \mu_i - \mumodeli - \DMUz )^2 / \sigmui^2  \label{eq:chisqHD} \\
    \sigmu^2 & = & 
         {\sigint}^2 + ({\sigmuzerr})^2  
                 \nonumber \\
            & + & C_{m_B,m_B} + \alpha^2 C_{x1,x1} + \beta^2 C_{c,c}  
                  \nonumber \\
           &  + &   2\alpha C_{m_B,x1} - 2\beta C_{m_B,c} - 2\alpha\beta C_{x1,c}~,
                  \nonumber \\
            \sigmuzerr & = & \left( \frac{5}{\ln(10)} \right) \frac{1+z}{ z(1+z/2) }    
                                            \sqrt{ \sigma_z^2 + (\sigvpec/c)^2 } ~.
                  \nonumber
\end{eqnarray}
The fitted parameters are $\alpha$, $\beta$, $\sigint$ and a distance offset 
($\DMUz$) in each redshift bin.
In the definition of the \unc\ term, $\sigmu$, the event index $i$ has been dropped.
$C$ is the fitted covariance matrix among the $\{m_B,x_1,c\}$ parameters,
$\sigint$ is the intrinsic scatter term, 
$\sigma_z$ is the redshift \unc, and
$\sigvpec$ is the peculiar velocity \unc.
The \Smu\ program can add an arbitrary $3\times 3$ intrinsic scatter
matrix ($\Cint$) in the calculation of $\sigmu$;  
here we use $\Cint_{m_B,m_B} = \sigint^2$ and set all other $\Cint$ terms to zero.
This choice can be interpreted as including only the coherent scatter model
(COH model in \S\ref{sec:sim})
in the distance \uncs, and ignoring intrinsic variations in color and stretch.

Rather than fitting for cosmological parameters  ($w,\OM$) 
appearing in the $\mumodeli$ term, 
the cosmology parameters are fixed to reference values ($\wref,\OMref$)
and  the $\DMUz$ are fit in $N_z$ redshift bins.
For typical fits the redshift bin size is $< 0.1$.  
For all event indices $i$ whose redshift lies in the same bin,  
$\DMUz$ is fixed to the same value. 
The key assumption here is that within each redshift bin,
the local shape of the HD is well described by the reference cosmological model,
and that the difference is characterized by an offset ($\DMUz$).
Thus instead of fitting for two cosmology parameters ($w,\OM$),  
\Smu\  returns $N_z$ fitted $\DMUz$ values
along with $\alpha$, $\beta$, and $\sigint$.
$\Mo$ in Eq.~\ref{eq:Tripp} is defined as the average of the $\DMUz$ values,
and each $\DMUz \to \DMUz - \Mo$.

Note that the $\chisqHD$ in Eq.~\ref{eq:chisqHD} has no \GN\ term, $-2\ln(\sigmu)$.
Minimizing this quantity will be referred to as the
 ``traditional $\chisqHD$'' method.

After obtaining the fitted $\DMUz$ from the \Smu\ program, 
cosmological parameters are obtained in a separate fit of 
$\murefz + \DMUz$ vs. redshift, where $\murefz$ are the 
distances computed from the reference cosmological parameters
($\wref,\OMref$) used in the \Smu\ fit.  
The \Smu\ program could in principle report an average distance 
modulus ($\mu_z \equiv \murefz + \DMUz$) in each redshift bin,
with no impact on subsequent cosmology fitting codes.
However, we prefer to report the fitted $\DMUz$ instead.

% ======================================================
% ======================================================
 \section{Simulations}
 \label{sec:sim}
% ======================================================
% ======================================================
%
We test  \acro\  on \SNANA\ simulations based on the 
cadence and observing conditions from the first DES-SN season.
The SN survey component of DES is described in \citet{DiffImg}, 
and the Dark Energy Camera (DECam) is described in \citet{DECAM2015}.
The survey consists of ten 3~deg$^2$ fields, observed roughly once per
week in each $griz$ passband.
Eight of these shallow fields are observed to an average depth of 23.5;
the other two deep fields are observed to an average depth of 24.5.
The first \DESSN\ season was used to build a library  consisting of 
sky noise, zero-point, and point-spread function (PSF) for each \obs\ in the ten fields. 
This library is used to simulate realistic light curves at random times and sky locations.
For each \obs, the simulated magnitude is converted into
a flux using the image zero-point and CCD gain. 
The simulated flux \unc\ is computed from the PSF, sky noise, and zero-point.

\begin{figure}[hb]
\centering
\epsscale{1.1} 
\plotone{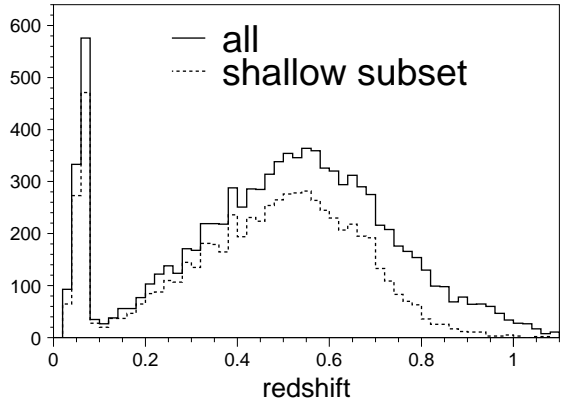}
  \caption{
    Simulated redshift distribution after selection requirements described in \S\ref{sec:NN}.
    The low-redshift ($z<0.1$) subset is $\sim 10$\% of the total.
  }
  \label{fig:simz}
\end{figure}

For SNe~Ia, the redshift-dependent volumetric rate ($R$) is taken from 
\citet{Dilday2008}, with $R(z) \propto (1+z)^{1.5}$.
For better statistical constraints on the $w$-precision of the \acro\ method, 
an artificial low-redshift ($z<0.08)$ sample is added,
which comprises $\sim 10$\% of the total sample as shown in Fig.~\ref{fig:simz}.
This \lowz\ sample is generated with the same $griz$ passbands and depth
as for the \DESSN\ sample, and is therefore an ideal anchor with minimal selection bias.

SN~Ia model magnitudes are generated 
from the \SALTII\ light-curve model in \citet{Guy2010}. Since this model is defined
in a limited rest-frame wavelength range (2800-7000~\AA), the model is undefined
for $i$ and $z$ bands at low redshift ($z<0.12$), 
and for the $g$ band at higher redshifts ($z>0.6$).
To avoid complications from missing bands in this analysis,
we have extrapolated the model into the ultra-violet and near-infrared 
regions (2200-9200~\AA).
While the extrapolated model is not calibrated at the percent-level precision needed
in a real cosmology analysis, it is sufficiently accurate to model Poisson noise
and to test fitting codes.
 
To match the observed Hubble residual dispersion,
we examine three different models of intrinsic scatter from \cite{K13}:
1) ``COH" model of 
    100\% coherent variation at all epochs and wavelengths,  
    with $\sigcoh = 0.13$~mag,
2) ``G10"  scatter from the \SALTII\ model in which 70\% of the contribution to the 
   Hubble residuals comes from coherent
   variation and  30\% comes from 
   chromatic variation, and
3) ``C11" model from \citet{C11}  in which 25\% of the contribution to the Hubble residuals 
   comes from  coherent  
   variation and 75\% comes from chromatic variation.

The COH model is not consistent with \obss\ 
because wavelength-dependent variations in the intrinsic
scatter have been observed in \citet{Guy2010,C11}. However,  
we include this model as a test because
the HD fitting model with constant $\sigint$ is similar to the COH model,
and thus one might naively expect that simulating the COH model 
would yield less biased results
compared to simulating the more realistic G10 and C11 models. 
The latter two models are broadband-variation models based on \obss, 
and were converted into spectral variation models in \citet{K13} 
so that intrinsic variation is simulated by varying the underlying \SALTII\ spectra.
These  models add intrinsic scatter without changing the average underlying SN~Ia model;
the average generated SN~Ia flux after intrinsic variations are applied
is the same as the \SALTII\ model flux with no intrinsic scatter.
The simulated values of the standardization parameters (Table~\ref{tb:SNpar})
are  $\aSIM = \agen $ and $\bSIM = \bgen $ for the COH and G10 scatter models;
for the C11 model $\bSIM=3.85$.
The larger $\bSIM$ value for the C11 model is taken from 
\citet{SK2016}, and is a result of the color population being more like 
a falling exponential compared to the G10 model. 
   
The underlying color and stretch populations are each described by
an asymmetric Gaussian distribution defined by three parameters.
Ignoring normalization factors, the color distribution is
\begin{eqnarray}
   P(c)  & \propto & \exp[-(c - \cbar)^2/ 2\sigminus^2 ] ~~~ (c \le \cbar) \\   
   P(c)  & \propto & \exp[-(c - \cbar)^2/ 2\sigplus^2 ] ~~~ (c > \cbar) \nonumber
\end{eqnarray}
where $\cbar$ is the value with maximum probability, and
$\sigminus$ and $\sigplus$ are the low- and high-sided Gaussian widths. 
A similar parameterization describes the stretch ($x_1$) distribution.
We use the parameters in the High-$z$ row in Table~1 of SK16,
which are shown here in Table~\ref{tb:SNpar} for each  
model of intrinsic scatter. 
Note that in a more realistic analysis, 
the redshift-dependent populations should be used.
\begin{table}[h!]
\caption{Simulation Parameters for SN~Ia Properties}
\begin{center} \begin{tabular}{ | l  | c | c  | c | }  \tableline  
               &    &  Color Params:            &  Stretch Params: \\
   Model  & $\aSIM$~$\bSIM$ &  $\cbar~~\sigminus~~\sigplus$  &  $\xbar~~\sigminus~~\sigplus$ \\ [1pt]
   \hline 
 COH\tablenotemark{a} 
           &  0.14~~3.20 & $-0.054~0.043~0.101$  &  $0.973~1.472~0.222$  \\
 G10   &  0.14~~3.20 & $-0.054~0.043~0.101$  &  $0.973~1.472~0.222$  \\
 C11    &  0.14~~3.85 &  $-0.099~0.003~0.119$  &  $0.964~1.467~0.235$  \\
\tableline \end{tabular} \end{center}   
 \tablenotetext{1}{SK16 did not evaluate population parameters for the COH model,
    so here we use the G10 parameters.}
\label{tb:SNpar} \end{table}

CC SN types II and Ibc are generated using the volumetric rate  from \citet{Bazin2009},
with a redshift dependence of $(1+z)^{3.6}$. 
Simulated light curves are generated from spectra that have been
mangled to match photometric observations of 42 CC light curves
as described in \citet[hereafter K10]{K10_SNPCC}.
The relative fraction  and peak luminosity function (LF) 
for each sub-type (II,Ib/c) are from \citet[hereafter L11]{Li2011}.
In K10, the simulation selects a random template spectrum and applies
a magnitude offset and random Gaussian scatter such that the 
generated LF (in $R$-band) has the same mean and variance as reported in 
Table~6 of L11.  
The  brightness distribution of the original CC light curves is preserved, 
and the LF from L11 is achieved with an additional random magnitude scatter.
Here we use the same technique for the Type-II SNe, but alter the procedure
for Type Ib to account for one anomalously bright event which overestimates 
the contribution from bright events. 
The brightness for each Ib template spectrum is adjusted to have the
mean LF brightness, and the full LF spread is simulated with random scatter
\citep{Jones2016}.

Since a \spec\ host galaxy redshift is required in this analysis, we use the
\spec\ matching efficiency estimated in \citet{DESSN2012}.\footnote{See Table 18
column with $m<24$ and $\kappa_{\rm Ia} = 0.5$.} 
This \eff\ drops to 0.75 at $z=0.5$, and drops to 0.50 at $z=0.95$. 

The following effects are available options in the \SNANA\ simulation,
but have been left out for this study: 
peculiar velocities, 
weak lensing, 
host-galaxy correlations, and
redshift-dependent SN properties (e.g., $\alpha$, $\beta$, population parameters).

Because the simulation generates realistic light-curve fluxes and \uncs,
there are no assumptions about the analytical form of
detection thresholds, and it properly simulates arbitrarily complex 
surevy-selection triggers based on the number of detections, 
as well as the distribution of detections
over nights and  passbands.
It also accounts for time-dependent variations
due to weather and instrumental effects. 
We  fit each simulated light curve with the same \SALTII\ model (and code)
used on data, and thus light-curve fitting biases are included, such as those found in SK16.

% ======================================================
% ======================================================
 \section{NN Method for Photometric Classification}
 \label{sec:NN}
% ======================================================
% ======================================================
%
Here we describe the  analysis to photometrically select SNe~Ia.
The overall strategy is to first apply selection cut-windows, or ``box cuts'', 
on several analysis variables to reduce the  CC/Ia fraction to $< 10$\%.
The second step is to apply the NN  method 
\citep[hereafter referred to as NN]{Sako2014}
to further reduce the CC \con\ and to determine a Type Ia \prob\ 
needed for the \acro\   \lh\  (\S\ref{sec:bcd}). 
Numerous machine-learning (ML) methods can be applied to photometric classification
as described in \cite{ML2016}.
Our choice of NN is arbitrary and adequate to test \acro;
we make no claim about which ML method is best.

After fitting the light curves with the \SALTII\ model,
the box cuts are as follows:
\begin{enumerate}
   \item at least one \obs\ with $\Trest<-2$~days, where $\Trest$ is the rest-frame
           epoch with respect to the epoch of peak brightness.
   \item at least one \obs\ with $\Trest > +10$~days.
   \item at least three bands have an \obs\ with a signal-to-noise ratio (S/N) above 5.
   \item redshift $z < \zmax$.
   \item $\vert x_1 \vert < 3 $  and $\vert c-0.1\vert <0.4$~.
   \item the \SALTII\ light-curve fit \prob\ ($\Pfit$), computed from the fit-$\chi^2$
            and number of degrees of freedom, satisfies $\Pfit>0.05$~. 
\end{enumerate}
The first four requirements ensure good light-curve quality,
while the last two requirements select SNIa-like light curves.
At this stage of the analysis, we transition to the more sophisticated 
NN method to further reduce CC background.

Our  NN analysis is based on the 3D  space of $\{z,x_1,c\}$,
where $z$ is a \spec\ host galaxy redshift and the latter two variables are 
from the \SALTII\ light-curve fit. 
For a given data event,\footnote{We use the term `data' here, even though it is a simulated
 data sample, since the NN procedure is the same with real data.}
the basic idea is to define a sphere centered at $\{z,x_1,c\}$, 
generate a  large simulated sample, and 
count the number of Type Ia and CC SNe that are found inside the sphere. 
The event is classified to be the SN type representing
the majority inside the sphere. 
A procedure called `NN training' determines the optimal size of the sphere
based on maximizing the product of the  \eff\ and purity.

More formally, for each event in the data sample,
the  NNs are  events from a large simulated training sample that satisfy 
a 3D distance constraint,
\begin{equation}
    d^2 =  \left[
    \frac{(z-z')^2}{\Dz^2}  + \frac{(c-c')^2}{\Dc^2}  + \frac{(x_1-x_1')^2}{\Dx^2}    
     \right]   < 1 ~,
     \label{eq:NNdist}
\end{equation}
where the primed quantities are from the simulated training sample.
The optimal distance-metric parameters ($\Dz,\Dc,\Dx$) are determined
from a training procedure that maximizes the product of the 
SN~Ia purity and the \eff. 
The final selection requirement is that for simulated neighbors satisfying Eq.~\ref{eq:NNdist},
more than half are true SNe~Ia with at least $1\sigma$ confidence.
Thus if there are 20 neighbors,  then at least 13 ($\sigma=2.1$) must be true SNe~Ia
to be classified as SN~Ia.   If only 12 ($\sigma=2.2$)  of the neighbors are true SNe~Ia,
then the $1\sigma$-above-half requirement fails and the event is classified as unknown.
The Type Ia \prob\ for each event, $\PIa$,  is defined as the fraction of 
NN training events (satisfying Eq.~\ref{eq:NNdist})  that are true SNe~Ia.

\begin{figure}[h!]
\centering
\epsscale{1.2} 
\plotone{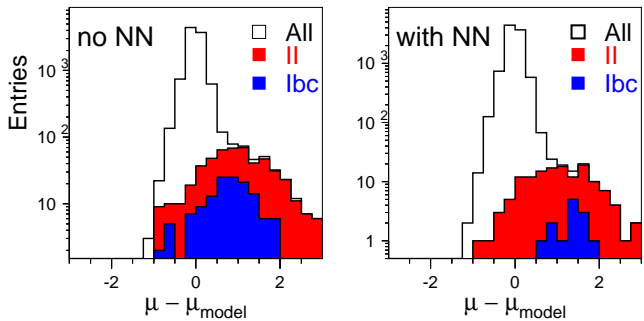}
  \caption{
   Distance modulus residual ($\mu-\mumodel$) for DES-SN simulation with $\RCC=1$:
   after box cuts (left) and after NN requirement (right).
   Legend indicates  contribution from all SN types (Ia+CC), and from types II and Ibc.
  }
  \label{fig:sim_mures}
\end{figure}

After passing the box cuts and NN requirement, there are a total of 
\NSIMDATA\ events in the simulated data sample, and 
\NSIMTRAIN\ events in each of the two training 
samples.\footnote{One of the training samples serves as an independent data sample
   to avoid statistical anomalies from self-training.}
With our nominal estimate of the CC rate, 
the CC {\con}\footnote{CC \con\ is defined as the fraction of events that are true CC.}
is $\CCfracBoxCuts$ after the selection requirements,
and drops to $\CCfracNN$ after the NN requirement. 
The corresponding SN~Ia loss from the NN requirement is $\SNIAineffNN$.
Since the CC \con\ in a photometrically identified sample is not yet known, 
we test the \acro\ method with two different CC rates.
Simulations are generated with our best estimate of the  
CC rate ($\RCC=1$), and again with  $3\times$ the CC rate ($\RCC=3$).  
$\RCC$ is defined here as the simulated CC rate divided by our
best estimate of the rate.
For each $\RCC$ value, the CC \con\ before and after
the NN requirement is shown in Table~\ref{tb:NN}.
Fig.~\ref{fig:sim_mures} shows the distance modulus residuals to
illustrate the \con\ before and after the NN requirement.

\begin{table}[h]
\caption{ CC Contamination Fraction and SN~Ia Efficiency vs. $\RCC$, \\ Before and After NN Requirement}
\begin{center} \begin{tabular}{ | c  | c  | c | c | } \tableline  
                        & CC/All  & CC/All     &  Eff(SNIa)\\
   $\RCC$         & no NN   & with NN   & with NN \\
   \hline
     1       & $\CCfracBoxCuts$    & $\CCfracNN$   & $\SNIAeffNN$ \\
     3       & $\CCfracBoxCutsx$  & $\CCfracNNx$  & $\SNIAeffNNx$ \\
\tableline  \end{tabular} \end{center}   
\label{tb:NN} \end{table}

A final caveat is that the NN training has been performed on the 
combined deep+shallow fields, but in principle a separate training
on deep and shallow sub-samples could be more optimal. 
However, since the same requirements are applied to the data and 
simulated samples,
the level of NN optimization has no impact on the \acro\ performance.

% ======================================================
% ======================================================
% \bigskip
 \section{ {\acro} Method}
 \label{sec:bcd}
% ======================================================
% ======================================================
%
The \acro\ method consists of two fitting stages. 
The first step is the \acro\ fit to determine $\DMUz$ in each redshift bin,
where $\DMUz$ is the distance-modulus offset with respect to an arbitrary 
cosmological model, which will be referred to as the reference cosmology. 
In addition to requiring light curve fit results ($x_0,x_1,c$) from the data,
the \acro\ fit also requires input from a detailed simulation
based on a rigorous characterization of the survey.
The second step is to fit $\DMUz$-vs-redshift in order to
determine the cosmological parameters. This step requires
no understanding of the survey, and hence can be performed
with a variety of fitting codes and priors.

% -----------------------------------------------
\subsection{Overview}
\label{subsec:method_overview}
% -----------------------------------------------
%
Using the BEAMS formalism, we form a \post\
consisting of Type Ia  and  CC \lhs.\footnote{See Eq.~2 in H12} 
The joint \post\  for $N$ events is
$\prior \times \LH$, 
where $\prior$ is a flat prior on the fitted parameters ($\theta$),
$\LH \equiv \Pi_{i=i}^N \LH_i$, 
and $\LH_i$ is the \acro\ \lh\ for each event $i$.
{\tt MINUIT} is used to minimize $-2\ln(\LH)$.
The \acro\  \lh\  for each event is 
\begin{eqnarray}
   \LH_i  & \equiv  &   \left[  \frac{\PIa^i}{\Ptot^i}  \DIa(z_i,\mu_i,\mumodeli )   \right]   \label{eq:L} \\
           & +  &   \left[ \frac{{\SCC}^i (1-\PIa^i)}{\Ptot^i}  \DCC(z_i,\mu_i,\mumodeli) \right]~,  \nonumber 
\end{eqnarray}
where $z_i$ is the redshift and $\mu_i$ is the distance modulus.
Dropping the event index $i$,
\begin{equation}
      \Ptot  =  \PIa + \SCC(1-\PIa)~,   \label{eq:Ptot}
 \end{equation}	
and $\DIa$ and $\DCC$ are the conditional \lhs\ for SN~Ia and CC, respectively,
\begin{eqnarray}
      \DIa & = &   \exp[-\chisqHD/2] / \sigmu\sqrt{2\pi}   ~~~~ \label{eq:DIa}  \\
      \DCC & = & {\rm map~from~simulation.}    \label{eq:DCC} 
\end{eqnarray}
$\DIa$ and $\DCC$ are  normalized such that at any redshift the 
distance modulus integrals satisfy  
\begin{equation}
   \int_{-\infty}^{+\infty} \DIa(z_i,\mu_i,x) dx  =  \int_{-\infty}^{+\infty} \DCC(z_i,\mu_i,x) dx = 1~,
   \label{eq:norm}
\end{equation}
and $\int \LH_i d\mu = 1$ for each event.
$\DIa$ (Eq.~\ref{eq:DIa}) depends on the traditional $\chisqHD$ term defined in 
Eq.~\ref{eq:chisqHD}, and includes the fitted $\DMUz$.
$\PIa$ is the NN \prob\ for Type Ia, and 
$(1-\PIa)$ is the NN \prob\ for CC. 
$\SCC$ is a fit parameter allowing for an arbitrary scale  of the CC \prob;
a correct CC simulation should result in $\SCC \sim 1$.
\footnote{Note that $\SCC \ne 1$ implies that $\Ptot \ne 1$ (Eq.~\ref{eq:Ptot}),  
but the overall normalization, $\int \LH_i d\mu = 1$, is always satisfied.} % end footnote
Although we use a specific case of NN \probs\ in the \lh, this method works
using \probs\ from any classification method.

The fitted parameters and flat prior ranges  are 
summarized in Table~\ref{tb:fitParDef}.
Compared to minimizing the traditional $\chisqHD$, the only additional  fit parameter 
in $\LH$ is $\SCC$. 

\vspace{-0.5cm}
\begin{table}[h]
\caption{\acro\ Fit Parameters}  \vspace{-0.5cm}
\begin{center} \begin{tabular}{ |  c  | c | }  \tableline  
                         & Flat Prior Range: \\
  Fit Parameter  & min, max \\
 \hline
 $\alpha$    &  0.02, 0.30 \\
 $\beta$      &  1.0, 6.0  \\
 $\SCC$      &  $-0.1$, 5.0  \\
 $\DMUz$   &  $-5.0$,  5.0 \\
\tableline \end{tabular} \end{center}   
\label{tb:fitParDef} 
\end{table} 

\vspace{-0.5cm}
We make two improvements with respect to the BEAMS analysis in H12:
1) the \SALTII\  fitted parameters ($m_B,x_1,c$) are bias-corrected based on a
simulation of SNe~Ia, and
2) the analytical functional form for $\DCC$ in H12 is replaced  by a simulated 
probability map to remove assumptions about the analytical form.
The sub-sections below describe these improvements in detail.

While the traditional $\chisqHD$ (Eq.~\ref{eq:chisqHD}) has no \GN\ term
to avoid biases, $\DIa$ (Eq.~\ref{eq:DIa})  must include this term
to properly normalize the Ia and CC probabilities. 
Thus the \GN\ term cannot be arbitrarily removed from the \acro\ \lh,
as has been done with the traditional $\chisqHD$ method.
The correct solution is to include bias corrections as described in 
\S\ref{subsec:biasCor}; 
this issue is examined further in \S\ref{subsec:wrongFits}.

We refer to the simulated data sample as `data', to clearly 
distinguish the data from a large independent  simulated sample used for 
bias corrections and the CC \lh. Thus the descriptions are valid when 
replacing a simulated data sample with real data.

Finally, since the shallow and deep-field samples differ in depth
by 1~mag, bias corrections and the CC \lh\  are determined 
separately for each sub-sample. This separation of the corrections
illustrates the more general principle of combining multiple samples
from different surveys.

% ----------------------------------------------------
\subsection{Bias-corrected Distance}
\label{subsec:biasCor}
% ----------------------------------------------------
%
For each data event, a bias-corrected distance is defined by replacing the 
fitted parameters  ($m_B,x_1,c$) in Eq.~\ref{eq:Tripp} 
with bias-corrected parameters,
\begin{eqnarray}
     \mucor & = & \mBcor + \alpha\xcor  - \beta\ccor - \Mo    \label{eq:mucor}  \\
                 & = & (m_B- \dmB) + \alpha(x_1 - \dx )  - \beta (c-\dc) - \Mo~~~   \nonumber  
\end{eqnarray}
where the star superscript indicates a bias-corrected quantity.
The bias corrections ($\dmB,\dx,\dc$) are determined from a large ``{\biasCor}" 
simulation with $\NSIMBIASCOR$ events after the requirements in \S\ref{sec:NN}.
For any given event, we cannot exactly determine the true parameter bias ($\delta$) 
because of variations caused by intrinsic scatter and measurement noise.
We therefore interpolate the bias ($\bar{\delta}$) in a 5D space 
of  $\{z,x_1,c, \alpha,\beta\}$.
The cell sizes are  $\{\CELLSIZEz, \CELLSIZEx, \CELLSIZEc\}$ 
for $\{z,x_1,c\}$, respectively.
The bias correction has a  weak dependence on $\alpha$ and $\beta$,
and is included by
generating these two parameters on a $2\times 2$ grid 
that extends well beyond current constraints,
and interpolating the bias within the \acro\ fit.
With current estimates of $\aSIM,\bSIM$  given in Table~\ref{tb:SNpar},
the two fixed $\alpha$ values are $\aSIM \pm 0.04$ and 
the two fixed $\beta$ values  are $\bSIM \pm 0.4$.  
The SN~Ia parameters $\alpha$ and $\beta$ are generated at discrete values
because there is no observational information about these distributions,
and thus we impose a flat prior in the \acro\ fit.
Each 3D sub-cell of $\{z,x_1,c\}$, however, includes a continuous 
distribution based on the  measured
SN rate vs. redshift and the  measured population of stretch and color.

Each $\bar\delta$-bias value ($\dmB,\dx,\dc$) is determined by linear interpolation 
in the 5D space. The bias and grid location in each 5D cell is the weighted average
among all \biasCor\ events in the cell, and each weight is $\sigmu^{-2}$.
A valid \biasCor\ cell requires at least \NPERCELLMIN\ events,
and at least \NCELLMIN\  valid cells are required for interpolation.
Data events with fewer than \NCELLMIN\ cells are rejected,
which reduces the CC \con\ as described in  \S\ref{sec:results-II}.
 
There is a subtle interpolation issue for events in the highest redshift bin
($1.05 < z < \zmax$). Recall that the bias value in each redshift bin is 
defined at the weighted-average grid location, $\zbin$,
and thus for $\zbin < z < \zmax$ the bias cannot be interpolated
if there are no events beyond $z>\zmax$.
While it is possible to extrapolate the bias for $z>\zbin$,
extrapolating a rapidly varying bias function could result in a large error.
To avoid extrapolating, the \biasCor\ simulation includes events with 
$\zmax<z<1.15$ to provide a reliable interpolation for events with 
$\zbin < z < \zmax$.

\begin{figure}[!]
\centering
\epsscale{.36} 
\plotone{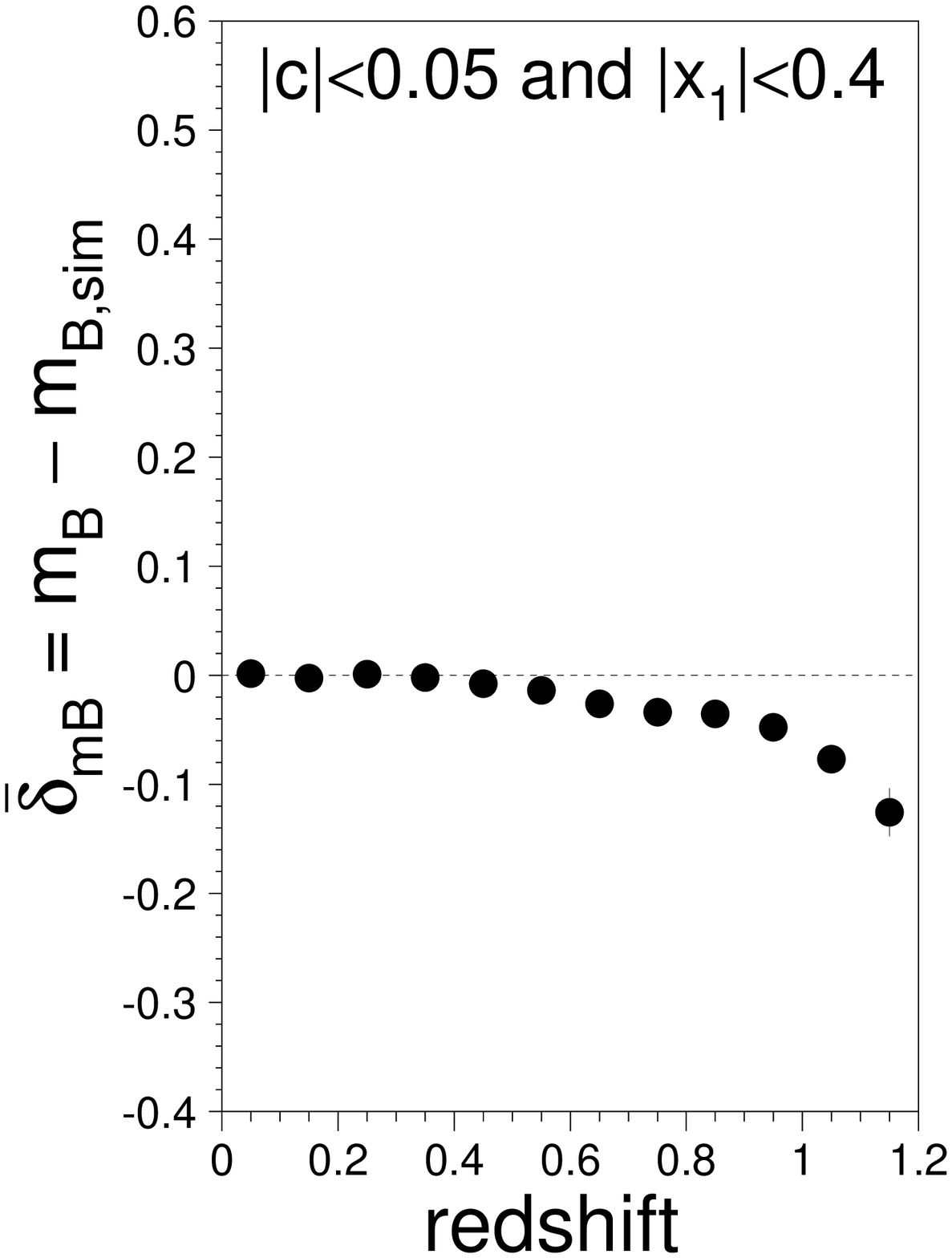}
\plotone{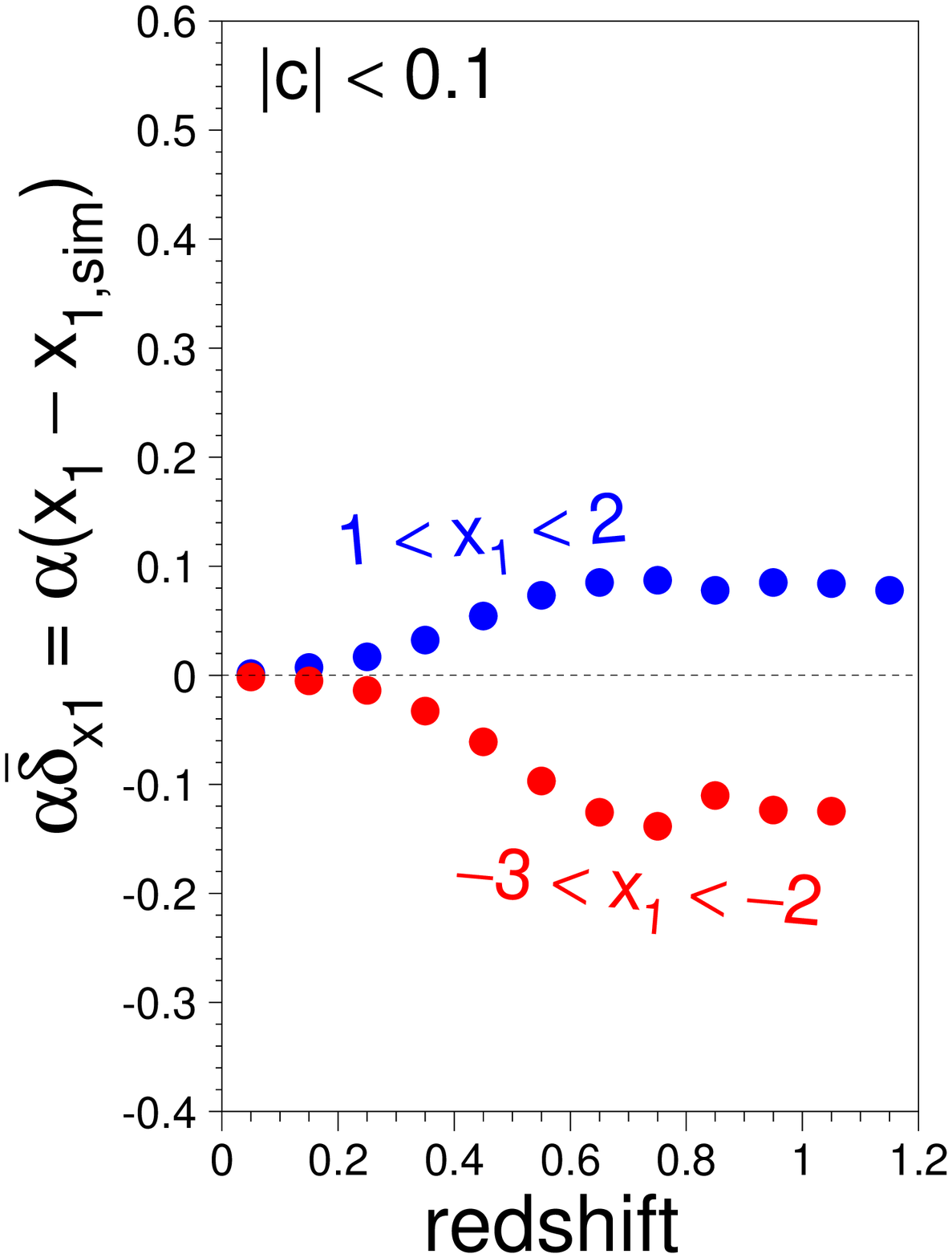}
\plotone{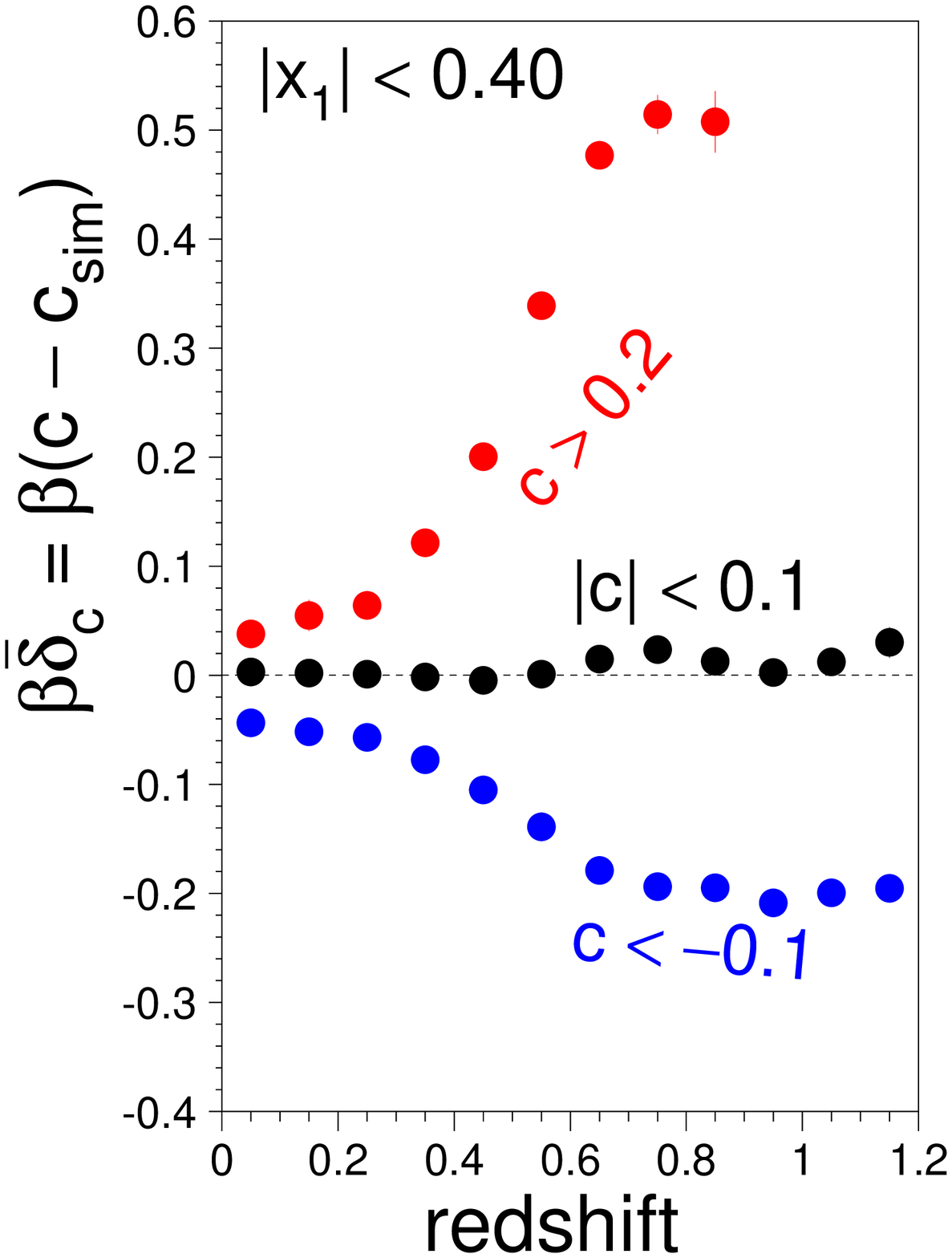}
  \caption{
   Bias corrections  $\dmB$, $\alpha\dx$, and $\beta\dc$ are shown
   as a function of redshift.
   The pre-factors $\alpha,\beta$ are used to show the bias in 
   distance-modulus magnitudes.
   The parameter selection ranges are shown on each panel. \\
  }
  \label{fig:biasCor}
\end{figure}

\begin{figure}
\centering
\epsscale{1.15} 
\plotone{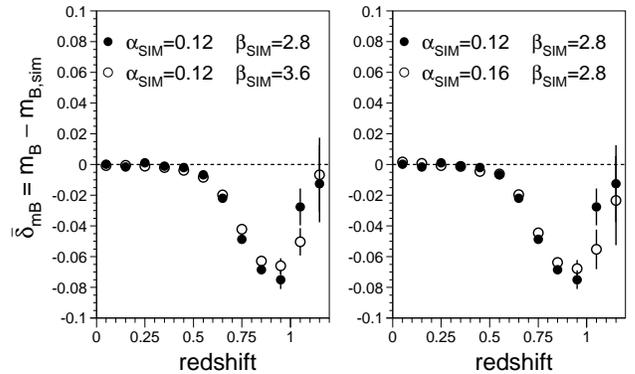}
  \caption{
   Bias correction $\dmB$ vs. redshift  for the entire shallow-field subset.
   Solid circles are the same in both panels. 
   Open circles show bias 
   with same $\aSIM$ and different $\bSIM$ (left),  and 
   with same $\bSIM$ and different $\aSIM$ (right).
   Note that the vertical scale is $\times 5$ smaller than in    Fig.~\ref{fig:biasCor}.
  }
  \label{fig:biasCor_ab}
\end{figure}

The bias corrections as a function of redshift are illustrated in Fig.~\ref{fig:biasCor}
for a few arbitrary stretch and color bins;  
the largest corrections are a few tenths of a mag for very blue and red colors.
These corrections are similar to the 3D $(z,x_1,c)$ corrections in SK16, 
except that here we account for  the dependence on $\alpha$ and $\beta$,
as illustrated in Fig.~\ref{fig:biasCor_ab}.

% ----------------------------------------------------
\medskip
\subsection{Bias-corrected Distance Uncertainty }
\label{subsec:biasCorErr}
% ----------------------------------------------------
%

\begin{figure*}[t!] 
\centering
\epsscale{.36} 
\plotone{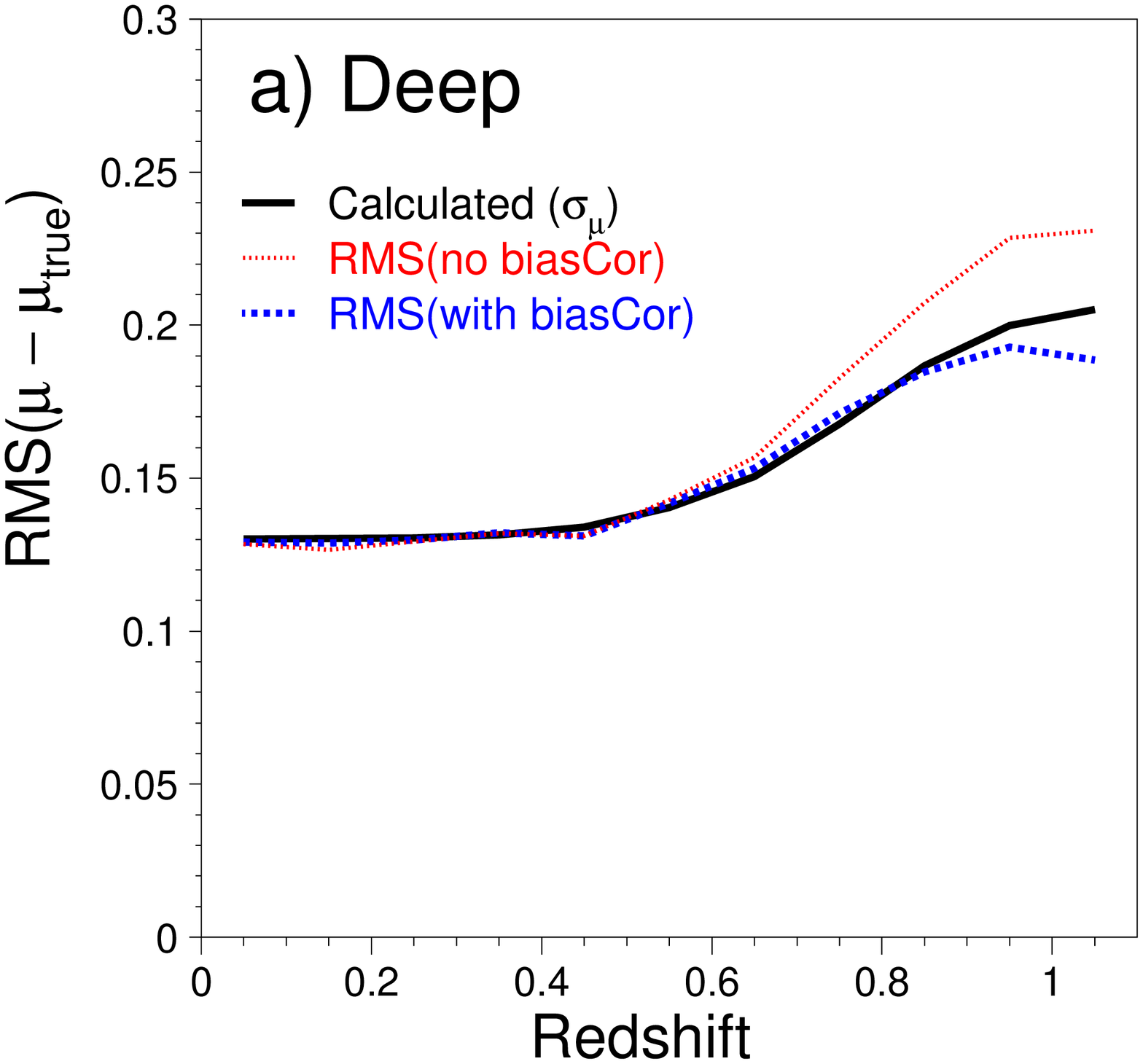}
\plotone{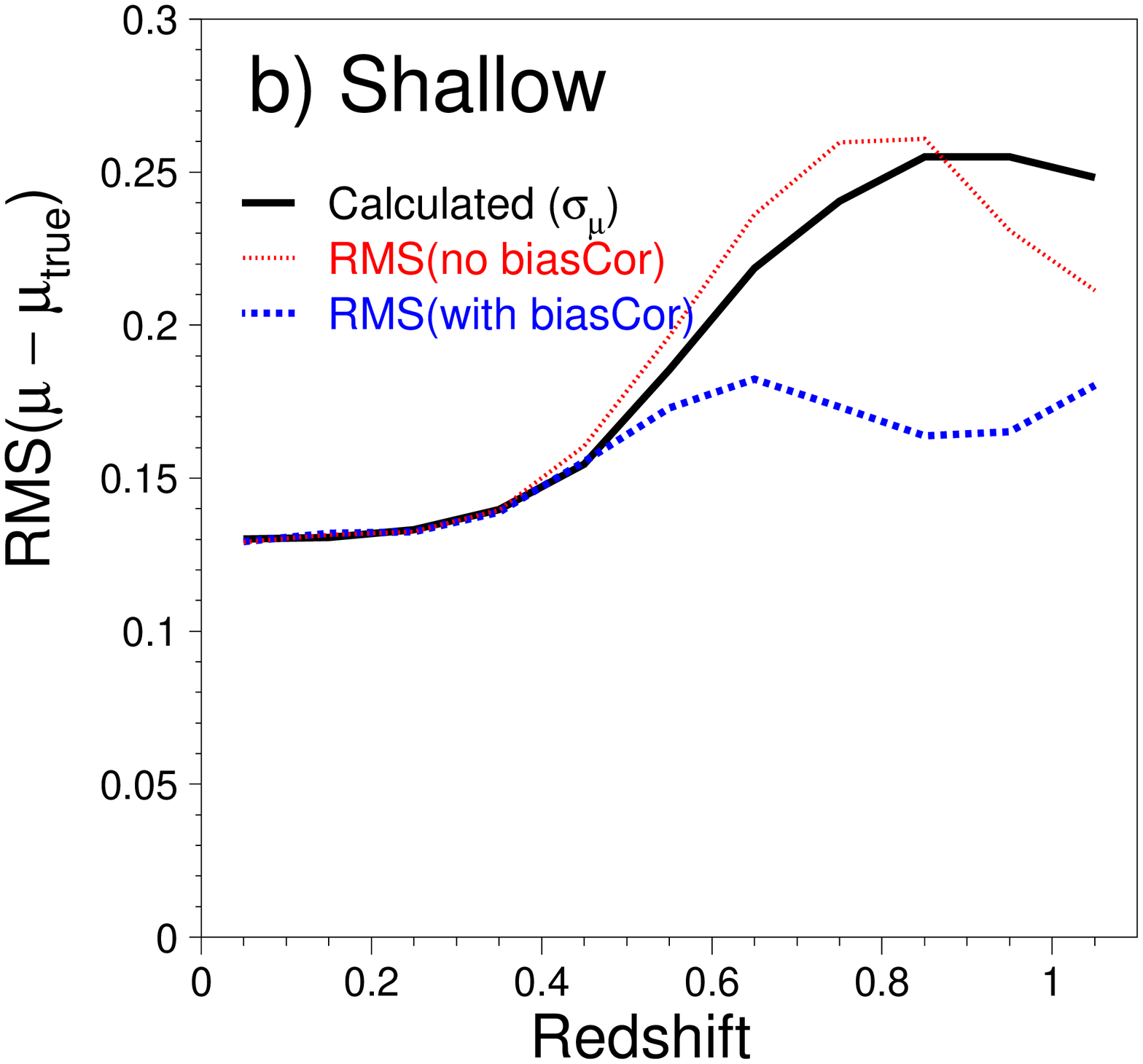}
\plotone{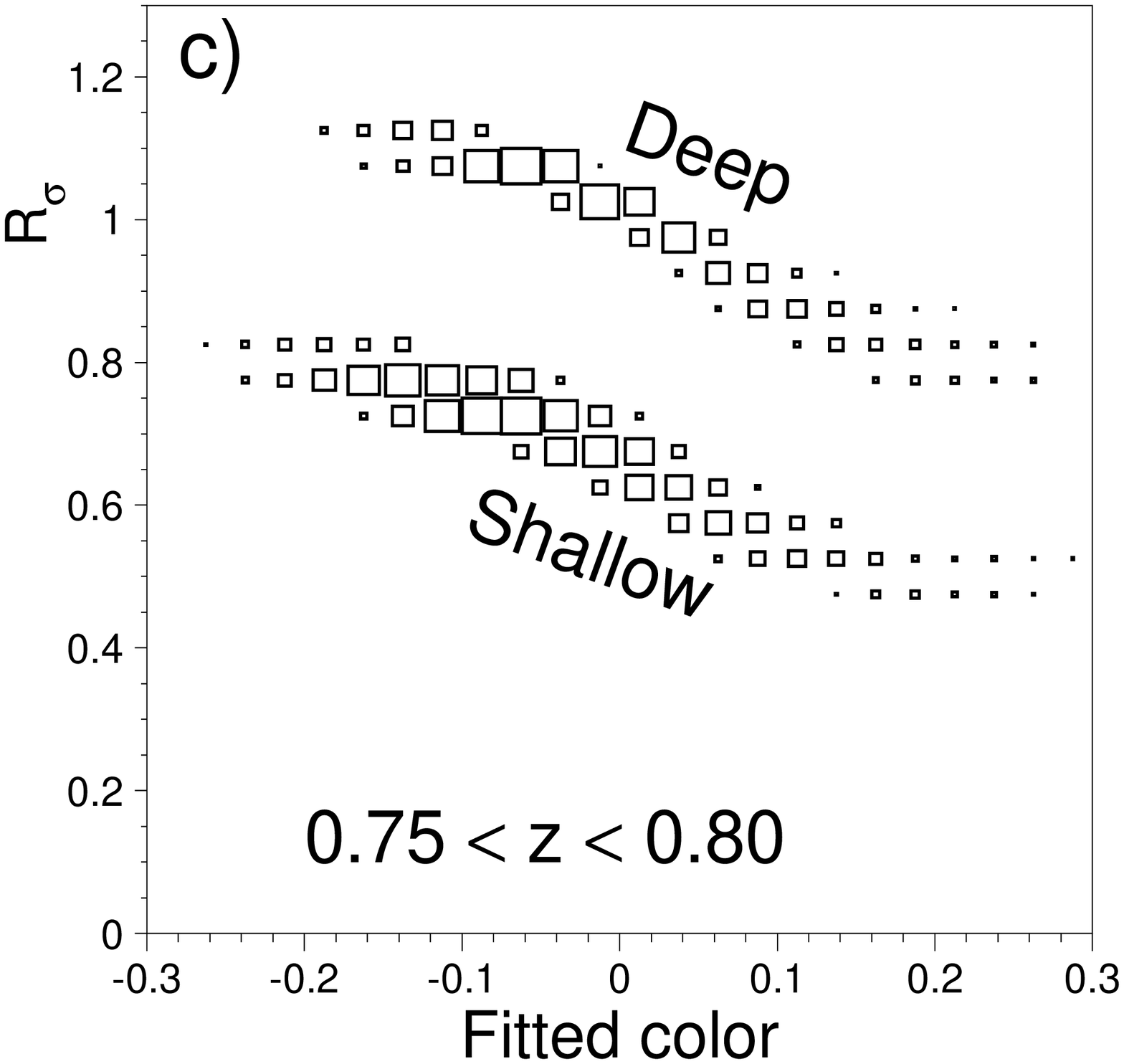}
  \caption{
    RMS of $\mu -\mutrue$ vs. redshift before bias corrections are applied (red-dotted),
    after bias corrections are applied (blue-dashed), 
    and calculated $\sigmu$ from Eq.~\ref{eq:chisqHD} (black line).
    Left panel (a) is for the 2 deep fields, and middle panel (b) is for the 
    8  shallow fields. 
    Right panel (c) shows the $\Rsig$ correction   (Eq.~\ref{eq:Rsig})  
    on $\sigmu$ vs. fitted color,
    in the redshift bin indicated on the panel. \\
  }
  \label{fig:muerr}
\end{figure*}

After applying bias corrections,  simulations show that
the Hubble residual scatter can be significantly reduced, and it can  
differ from the calculated \unc\ ($\sigmu$)  in Eq.~\ref{eq:chisqHD}. 
This effect is shown in Fig.~\ref{fig:muerr}a,b for the deep- and shallow-field subsamples, 
respectively, 
for Hubble residuals with respect to the true distance $\mutrue$. 
In the deep fields (Fig.~\ref{fig:muerr}a), bias corrections result in a $\sim 20$\% 
reduction in the Hubble scatter at the highest redshifts, 
while in the shallow fields (Fig.~\ref{fig:muerr}b) the reduction reaches $\sim 40$\%.
The calculated \unc\ ($\sigmu$) is in good agreement with the bias-corrected Hubble scatter
in the deep sample, but $\sigmu$ is much too large in the shallow sample.

To ensure a robust estimate of the  distance \unc\ in the \lh, 
we applied a 2D $\sigmu$ correction 
as a function of redshift and color.  
We found little $\sigmu$ dependence on the stretch ($x_1$) parameter
and therefore did not include an $x_1$ correction. 
The color dependence is illustrated in Fig.~\ref{fig:muerr}c.
The redshift binning is the
same as for the bias corrections, but there are only 3 color bins in order
to maintain better bin statistics for measuring the rms.
The form of the correction is
\begin{eqnarray}
    \sigmuData & \to & \sigmuData \times \Rsig  \\
          \label{eq:sigmuCor1}
      \Rsig & \equiv &  {\rm rms(with~biasCor)} / \sigmuSim~,
         \label{eq:Rsig}
\end{eqnarray}
where $\Rsig$ is determined before the \acro\ fit 
from the same \biasCor\  simulation used
to determine the bias corrections;
it is the ratio of blue and black curves in Fig.~\ref{fig:muerr}a,b.
The simulated intrinsic scatter term ($\sigint$) is 
determined from the subset with signal-to-noise ratio $S/N > 60$, 
and it is important to verify that $\Rsig \simeq 1$ at low redshifts with high $S/N$.  
$\Rsig(z)$ is determined at each $\alpha$-$\beta$ grid point,
as well as for deep and shallow subsamples.
During the minimization, $\Rsig$ is interpolated as a function of 
$\alpha,\beta,z,c$.
After the \acro\ fit, we verify that  rms$[(\mu-\mumodel)/\sigmu]$ is close to 1 at all redshifts.
A similar test on real data would check the validity of the $\Rsig$ correction.

An unexpected artifact of the bias correction is that the Hubble scatter at high 
redshift is somewhat smaller in the shallow-field subsample than in the deep fields,
which contradicts our naive expectation that the scatter should be smaller in the 
deep fields (compare blue-dashed curves in Figs.~\ref{fig:muerr}a,b). 
The reason for this paradox is that at high redshifts a narrower range of brightness is
observed in the shallow fields, resulting in smaller scatter after bias corrections.
We do not, however, achieve better measurements in the shallow fields for two reasons.
First, there are more events per square degree and redshift bin in the deep fields. 
Second, the shallow-field bias corrections are more sensitive to the cosmological parameters 
assumed in the \biasCor\ simulation, resulting in a larger systematic \unc\ 
(\S\ref{subsec:wrefBias}).

% ---------------------------------------------------------------------
%\clearpage
\subsection{CC Probability Distribution}
\label{subsec:CCprob}
In H12, $\DCC$ in Eq.~\ref{eq:L} has the same functional form as $\DIa$,
except that $\mumodel\to \mumodel + \Upsilon(z)$ where $\Upsilon(z)$ 
is a polynomial function of redshift with coefficients as additional fit parameters. 
The error term ($\sigmu$) has the same functional form as in the SN~Ia \lh, 
but has an independent $\sigint$ term.
The Gaussian form of $\DCC$ trivially satisfies
the normalization condition (Eq.~\ref{eq:norm}), but may not be a sufficiently
accurate model. 

\begin{figure}[h]
\centering
\epsscale{1.1} 
\plotone{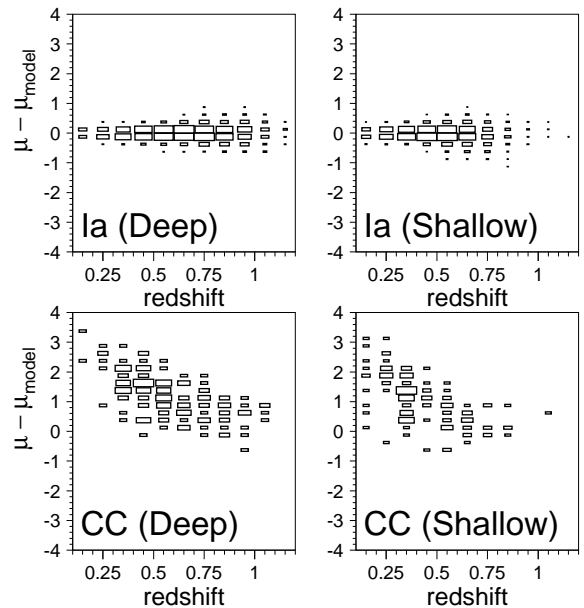}
  \caption{
   Distribution of distance residual vs. redshift, 
   generated from the simulation for SN~Ia (top) and CC SNe (bottom).
   Deep \& shallow subsamples are shown separately in the left  and right panels.
   Box cuts plus the NN requirement have been applied. 
   For the CC \lh, the simulated CC distribution in each redshift bin is 
   replaced with a Gaussian. 
   The SNIa distribution is shown for comparison, but the analytic SNIa \lh\  is used instead.	
  }
  \label{fig:CCmap}
\end{figure}

Here we replace $\Upsilon(z)$   with a simulated
map shown in the lower panels of Fig.~\ref{fig:CCmap}. 
As described in \S\ref{sec:sim}, the simulation is based
on 42 CC templates and a library of  observing conditions,
and makes no analytical assumptions about the form of the resulting
Hubble residuals.  
Additional motivation for using a simulated CC map is provided in
\citet{Jones2016}, where they show excellent agreement between
their \PS\ photometrically identified SN~Ia data sample and the Ia+CC simulation.
The normalization constraint  in Eq.~\ref{eq:norm} is imposed numerically.
The CC map in Fig.~\ref{fig:CCmap} clearly has discontinuities in the
derivatives, which can cause problems with {\tt MINUIT} minimization.
To alleviate such fitting issues, the mean and rms of the CC map
in each redshift bin are used to define a Gaussian function.
Improved simulations in the future may suggest a more complex
function such as an asymmetric Gaussian.

% ---------------------------------------------------------------------
% \clearpage
\subsection{ Determination of $\sigint$ }
\label{subsec:sigint}

Using the traditional $\chisqHD$ method (Eq.~\ref{eq:chisqHD})
for \specy\ confirmed SN~Ia samples, 
$\sigint$ is defined such that $\chisqHD = \Ndof$, where $\Ndof$ 
is the number of degrees of freedom in the fit. 
To maintain a similar definition of $\sigint$ in the \acro\ \lh,
we impose an ad hoc weighted $\chi^2$ constraint,
\begin{eqnarray}
    \left\{ \sum_i  \Wi  ( \mu_i - \mumodeli - \DMUz )^2 / \sigmui^2  \right\} & = &  \\
        \left\{ \sum_i \Wi \right\}  - \Nfitpar ,  & &  \nonumber
    \label{eq:Wi}
\end{eqnarray}
where $\Wi = \LH_{{\rm Ia},i}/\LH_i$ is the normalized SN~Ia \lh\ for event $i$,
$\LH_{{\rm Ia},i}$ is the first term in brackets in Eq.~\ref{eq:L},
and $\Nfitpar$ is the number of fitted parameters.
Note that we have defined $\sigint$  only for the SN~Ia \lh,
and there is no corresponding term in the CC \lh.
In the \Smu\ program, $\sigint$ is evaluated iteratively rather than
as a fit constraint. Convergence of Eq.~\ref{eq:Wi} to within 1\% 
typically requires 2-3 fit iterations.

To compare $\sigint$ among different subsamples, the \unc\ on this
variance term is needed.
Since our \acro\ fitting procedure does not return a  $\sigint$ \unc,\footnote{
We note that BHM methods determine the $\sigint$ \unc\ as part of the fit, 
and \citet{JLA} used a restricted log-\lh\ technique to include \uncs\ in their 
measurement of $\sigint$ as a function of redshift (see their REML discussion).
} % end footnote
fitting an ensemble of  simulations is used to measure the rms spread in $\sigint$, 
and this rms is interpreted as the \unc. 
The fitted $\sigint$ value itself provides another crosscheck because
we expect these values to be the same from a \acro\ fit to the data and simulation;
a discrepancy would point to an error in modeling the intrinsic scatter in the simulation.

% ---------------------------------------------------------------------
\subsection{ Fit for Cosmological Parameters }
\label{subsec:cosmoFit}
The \acro\ method consists of two fitting stages.
The first step is to determine $\DMUz$ by 
maximizing the \acro\ \post\ \prob\  in Eq.~\ref{eq:L}.
The second step, determining the cosmological parameters, is described here.
For a \specy\ confirmed SN~Ia sample, one can perform the cosmology fit
with either the distances for each SN, or with the $\DMUz$. 
However, for a photometrically identified sample with CC \con,
the $\DMUz$ are properly corrected while the individual distances are not.

As described in \S\ref{sec:SALT2}, the result of the \acro\ fit is a set of 
distance offsets vs. redshift bin, $\DMUz$, using a reference set of cosmological 
parameters ($\wref,\OMref$) that are fixed in the \acro\ fit. 
The final fitted values ($\wfit,\OMfit$)  are determined by minimizing 
\begin{equation}
     \chisqDelta =    \vecD {\COV}^{-1} \vecD~,
     \label{eq:chisqDelta}
\end{equation}
%  NBINz
where $\vecD \equiv \DMUz + \murefz - \mumodelz$.  
In each of the \NBINz\ redshift bins, $\DMUz$
is the {\acro}-fitted distance offset,  
$\murefz$ is the distance modulus computed from the
reference parameters $\wref$ and $\OMref$, and
$\mumodelz$ is the  distance modulus computed from the
floated parameters $w$ and $\OM$.  Within each redshift bin,
$\murefz$ and $\mumodelz$ are computed at the weighted-average
redshift, where each weight is $\sigmu^{-2}$.

$\COV$ is the total covariance matrix, including both statistical and systematics terms.
The size of $\COV$ is $\Nz^2 = \NBINz^2 = 484$, and this size is fixed
regardless of the size of the data sample. 
For our {\acro}-fit tests we include only the  diagonal terms with statistical \unc\ $\sigmuz^2$.
The off-diagonal reduced covariances are small (few percent) and ignored,
and we do not include systematic \uncs.

With diagonal $\COV$,
$\chisqDelta$ is very similar to $\chisqHD$ in Eq.~\ref{eq:chisqHD},
except that the sum over individual events is replaced with 
a sum over bin-averaged distances in redshift bins.
The \Smu\ program, which implements the \acro\ fit and produces the $\DMUz$, 
can also be used to minimize Eq.~\ref{eq:chisqDelta} with a suitable change
in fitting options. However, it is  advantageous to use
more specialized cosmology fitting programs that include 
priors from other constraints or more diverse cosmological models.

To minimize the \unc\ on measuring $w$-biases,
we impose a strong and unrealistic Gaussian prior on the matter density,
$\OM= \OMref \pm 0.0001$, and a flat prior on $w$ ($-1.5 < w < -0.5$).
Fits on data should use a more realistic $\OM$ prior.
To illustrate the \unc\ reduction, we have run crosscheck fits with
a weaker Gaussian prior of $\OMref \pm 0.02$; the $w$-bias results are 
consistent with the strong prior, but the \uncs\ are 2-3 times larger.

% ======================================================
% ======================================================
% \clearpage
 \section{Results-I:  Tests with Simplified Simulations }
 \label{sec:results-I}
% ======================================================
% ======================================================

We begin with a simplified  test of the \acro\ method such that
any observed bias in the cosmology or nuisance parameters would 
point to a  problem with the method. 
The simplifications in the simulated data and \biasCor\ samples are
1) exclude CC SNe,
2) generate  coherent intrinsic scatter model  
   (COH model in \S\ref{sec:sim})
so that the $\sigint$ term in the SN~Ia \lh\ exactly matches
the generated model of intrinsic scatter,
and
3) remove Galactic reddening and peculiar velocities.
In the light-curve fitting,  the \SALTII\ model \uncs\ are set to zero 
since there is no color variation in the intrinsic scatter model.
Since there are no CC events, NN training is skipped and $\SCC=0$ 
in  the \acro\ \lh\ (Eq.~\ref{eq:L}).
In spite of these simplifications, the selection biases are still present
and must be accounted for to obtain unbiased results.

\begin{figure}[h]
\centering
\epsscale{1.15} 
\plotone{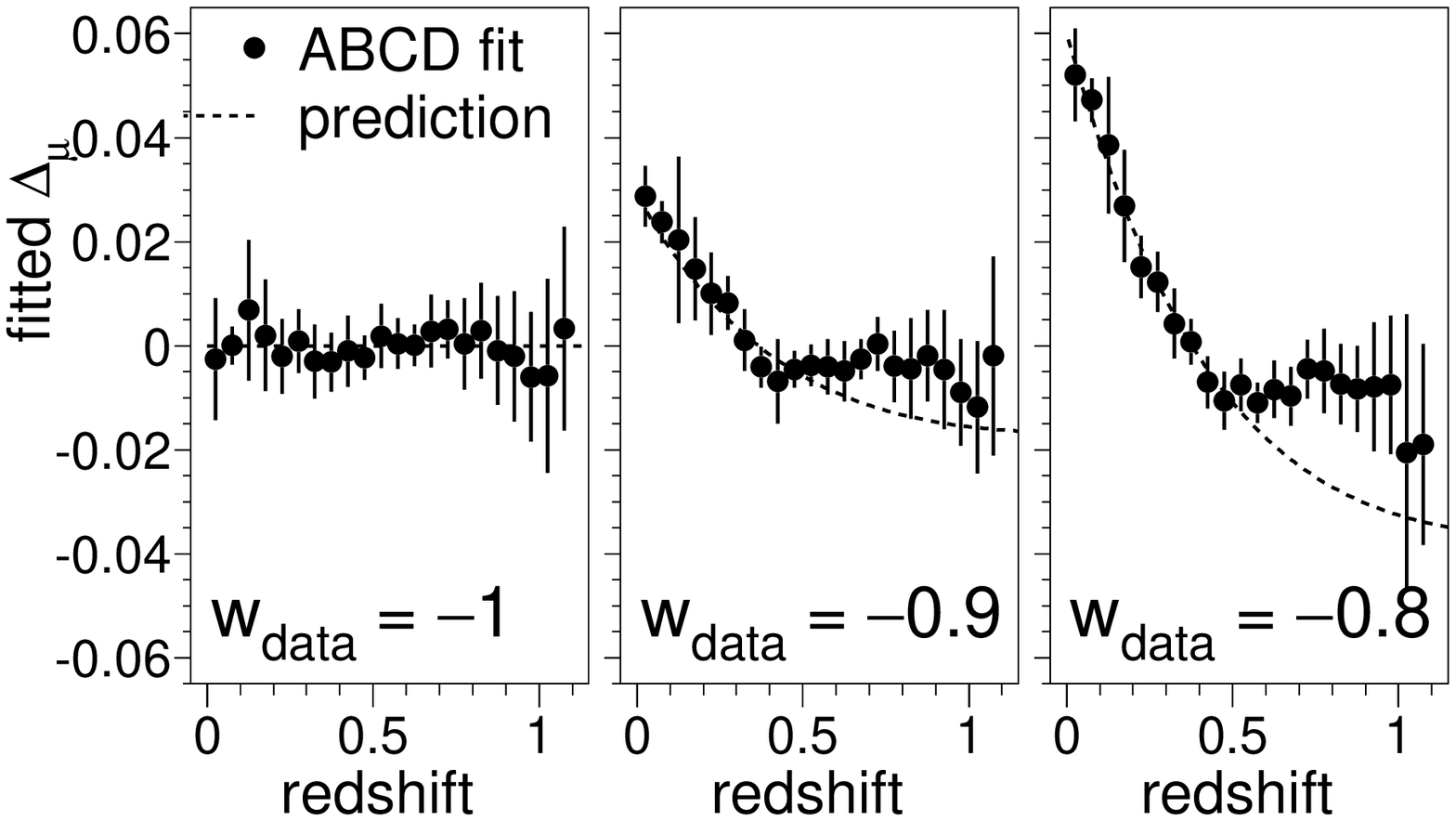}
  \caption{
    Average {\acro}-fitted value of $\DMUz$ vs. redshift, 
    and error bars show the rms spread among the \NSAMPLE\ samples.
    The simulated $\wdata$ is indicated on each panel: 
    	$-1,-0.9,-0.8$ in the three panels, respectively.
	The same \biasCor\ sample, with $\wref=-1$, is used in all three \acro\ fits.
	The dashed curve in each panel shows the prediction based on
	the $\Lambda$CDM model.   
    The discrepancies between the \acro\ fit
	and prediction (two right panels) is due to the difference between
	$\wdata$ and $\wref$. 
  }
  \label{fig:wfit_test}
\end{figure}

To reach a sensitivity to a \acro-induced $w$-bias below $0.01$,  
we simulate \NSAMPLE\ independent data samples, each with
$\sim\NSIMDATA$  events satisfying the box cuts and NN requirements 
in \S\ref{sec:NN}.
The \biasCor\ sample has $\sim\NSIMBIASCOR$ events,
and is used to correct all $\NSAMPLE$ data samples.
Distance moduli in all simulations are computed from a
cosmological model with a flat universe and $\OMref = \OMrefVal$.
We set $\wref=-1$ for the NN training (\S\ref{sec:NN}),  
\biasCor\ sample, and \acro\ \lh.
For the data samples we set the true $w$-value, $\wdata = -1$,
and also test with other values to check the
\acro\ method when the $\wref \ne \wdata$.
The $w$-bias is defined as
\begin{equation}
   w{\text-}{\rm bias} \equiv \wfit - \wdata~,
    \label{eq:wbias_def}
\end{equation}
where $\wfit$ is described in \S\ref{subsec:cosmoFit}.
Fig.~\ref{fig:wfit_test} shows the fitted $\DMUz$, 
averaged over the \NSAMPLE\ samples, vs. redshift.
In the left panel, $\wdata = -1$ and the fitted $\DMUz$ agree well with 
the expected curve (dashed line through zero).
In the next two panels of Fig.~\ref{fig:wfit_test}, 
$\wdata = -0.9$ and $-0.8$, which differs from $\wref=-1$ in the \acro\ fit.
The resulting $\DMUz$ show a significant redshift dependence,
and is expected as shown by the dashed curve. 
At higher redshifts, however, there is a notable discrepancy 
between the fitted $\DMUz$ and the prediction,
and this discrepancy is due to an incorrect bias correction
induced by a slightly incorrect cosmology model.
The maximum $\DMUz$ discrepancy is roughly 0.01~mag per
0.1 difference between $\wdata$ and $\wref$.
The resulting $w$-bias is quantified below in \S\ref{subsec:wrefBias},
and a potential solution is discussed in \S\ref{sec:fin}.

\begin{figure}[h]
\centering
\epsscale{1.1} 
\plotone{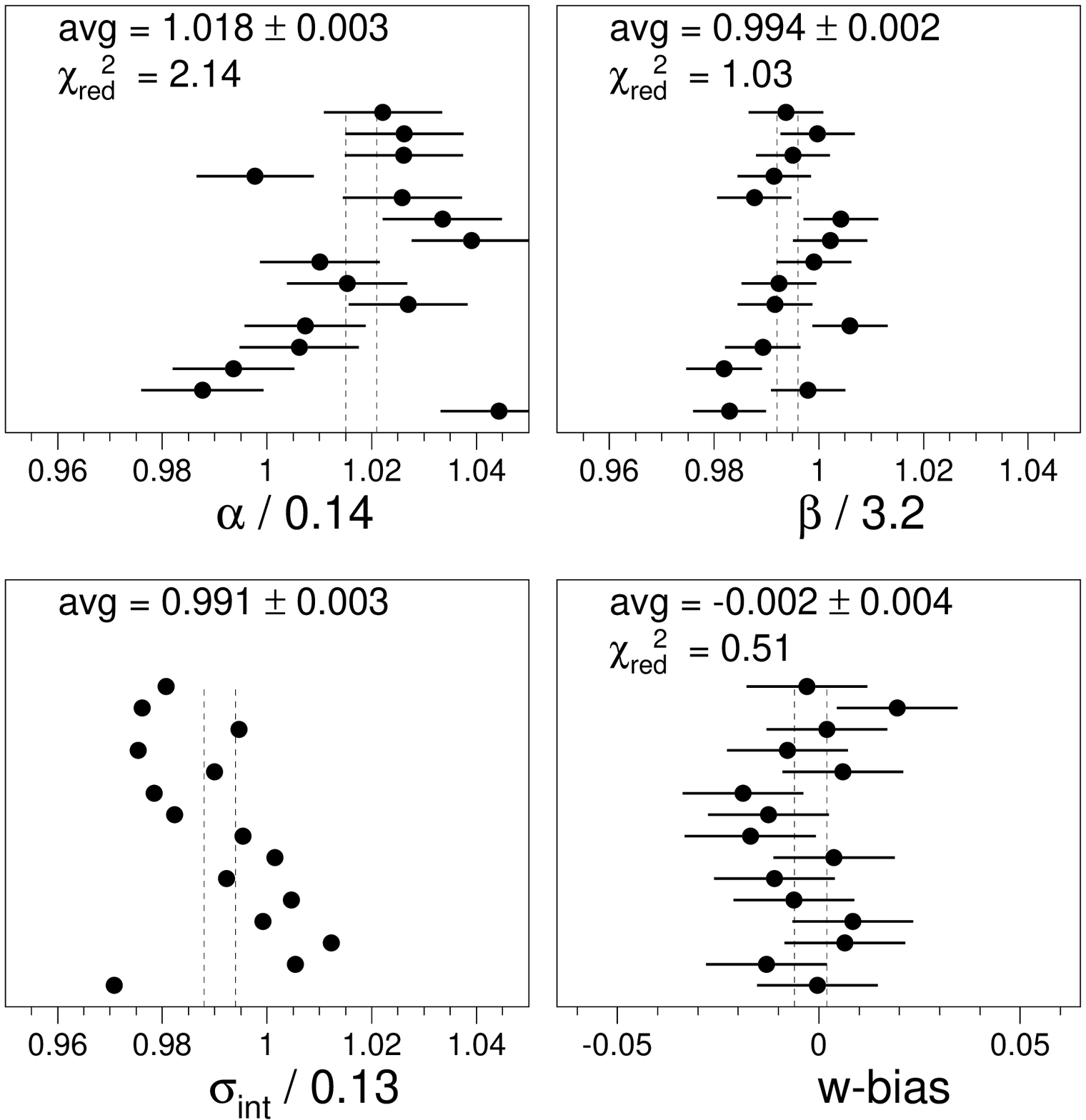}
  \caption{
   Results of \NSAMPLE\ simulated data samples fit with the SN~Ia \lh,
   and $\wdata = \wref = -1$.   
   First three plots show the ratio of the best-fit nuisance parameter 
   ($\alpha,\beta,\sigint$) to the true generated value.  Last plot shows
   the bias on $\wfit$ (w+1). 
   The ``avg" on each panel shows the weighted average
   and weighted \unc, except for the $\sigint$ panel where the \unc\ 
   is  ${\rm rms}/\sqrt{15}$. 
   $\chisqred$ in each panel is the reduced $\chi^2$ determined from the 
   \NSAMPLE\ independent fit-parameter values, their \uncs, and the weighted average.
   Vertical dashed lines show avg $\pm$ weighted \unc.  
  }
  \label{fig:results1_w1p0}
\end{figure}

For $\wdata=-1$, the $\acro$-fitted parameters for the \NSAMPLE\ samples
are illustrated in Fig.~\ref{fig:results1_w1p0} using an 
ideogram.\footnote{http://pdg.lbl.gov/2015/reviews/rpp2015-rev-rpp-intro.pdf}
The parameters $\alpha$, $\beta$, and $\sigint$ are shown as 
ratios with respect to the true input values, with the weighted-average ratio
printed at the top of each plot.
To evaluate the fitted \uncs, $\chisqred$ shows the reduced $\chi^2$
of the \NSAMPLE\ independent fits for each parameter;
good error estimates  result in 
$\chisqred \simeq 1$.\footnote{Be aware that $\chisqred$ is {\it not} computed from the \acro\ \lhs.} % end footnote
The $\alpha$ bias is almost 2\%, while $\beta$ and $\sigint$
are recovered to within $\sim 1$\%.
The $w$-bias is $\WBIAS$ (lower right panel of Fig.~\ref{fig:results1_w1p0}), 
and  $\chisqred \simeq \WCHISQRED$ suggests that the fitted 
$w$-\uncs\ may be overestimated,
although the \prob\ for such a low $\chisqred$ with correct \uncs\ is 6.5\%.

% ------------------------------------------------------------
\subsection{ Cosmology-dependent Bias in \biasCor\ Simulation} 
\label{subsec:wrefBias}
% ------------------------------------------------------------

Here we illustrate a subtle bias from using incorrect cosmology parameters
in the \biasCor\ simulation. 
For this test, $\wref$ is fixed to $-1$ in the \biasCor\ simulation,
while $\wdata$ takes on different values in the data samples.
For the three $\wdata$ values shown in Fig.~\ref{fig:wfit_test},
the {\acro} fit results and $w$-bias are shown in Table~\ref{tb:fitResults1}. 
The $\alpha$ bias persists at the 1-2\% level, with large variations in
 $\chisqred$ (0.4-2).
The $\beta$ parameter is measured to within 1\%, with a much smaller
spread in $\chisqred$ values.

\begin{table*}[t!]
\caption{ \acro\ Fit Results and $w$-Bias Averaged over $\NSAMPLE$ 
   Ideal DES-SNIa  Samples without CC Contamination \tablenotemark{a,b} }
\begin{center} \begin{tabular}{ | l |  cc | cc | c | rc |  }   \tableline  
 $\wdata$ & $\alpha/\agen$  &  $\chisqred$ & $\beta/\bgen$  &  $\chisqred$ & $\sigint/\sigCOH$  &
      $w$-bias\tablenotemark{d} & $\chisqred$  \\ [2pt]
   \hline\hline 
%
% from table_summary_w1p0_muopt000.tex (Nov 27 2016)
    $-1$\tablenotemark{c}  & 1.018(3) &  2.1 &   0.994(2) &  1 &   0.991(3) &   $ -0.003(4) $ & 0.5  \\
%
% from table_summary_w0p9_muopt000.tex (Nov 27 2016)
 $-0.9$  & 1.016(3) &  0.4 &   1.000(2) &  0.9 &   0.999(3) &   $ -0.017(4) $ & 0.4  \\
%
% from table_summary_w0p8_muopt000.tex (Nov 27 2016)
 $-0.8$  & 1.020(3) &  1.4 &   1.006(2) &  0.7 &   1.013(3) &   $ -0.044(4) $ & 0.8  \\
\tableline  \end{tabular} \end{center} 
\tablenotetext{1}{~Each value is the weighted average, and the value in () 
    is the weighted \unc\ in the last digit.} 
\tablenotetext{2}{$\chisqred$ is the reduced $\chi^2$ for the average among the \NSAMPLE\ samples.}
\tablenotetext{3}{~Results for $\wdata=-1$ are also shown in Fig.~\ref{fig:results1_w1p0}.}
\tablenotetext{4}{~$\wref=-1$ in each \biasCor\ sample; 
      $w$-bias can be reduced by iterating with $\wref\to\wfit$.} 
\label{tb:fitResults1} \end{table*}

Table~\ref{tb:fitResults1} shows a $w$-bias induced by the {\biasCor} dependence on 
cosmological parameters,
and this bias increases nonlinearly as $\vert\wdata-\wref\vert$ increases. 
Using the first two rows of Table~\ref{tb:fitResults1} to estimate the local derivative,
the $w$-bias is approximately given by $(\wref-\wdata)/7$.
This bias can be reduced with an iterative procedure in which 
$\wref$ is updated with the previous $\wfit$ value, but the proximity
of $\wref$ to the true value is limited by the total statistical+systematic \unc, $\sigw$.
Since $\wref$ has the same \unc\ as $\wfit$, there is an additional and irreducible
$w$-\unc\ of $\sim\sigwAdd$ induced by the {\biasCor} dependence on 
cosmological parameters.
The corresponding \uncs\ in the deep-  and shallow-field subsamples are
$\sigwAddDeep$ and $\sigwAddShallow$, respectively, 
and illustrates that this $w$-\unc\   depends on the SN sample.

To gain further insight, we repeated the \acro\ method with $\Rsig=1$ (Eq.~\ref{eq:Rsig}).
This test corresponds to using the black curves in Figs.~\ref{fig:muerr}a,b, 
instead of the blue-dashed curves.
We find that the bias is reduced by almost a factor of 2. 
With $\Rsig=1$ the distance \uncs\ are larger at higher redshift,
while the resulting $\sigint$ is 7\% smaller in order to satisfy the
$\chisqHD$ constraint in \S\ref{subsec:sigint}.   
This change in \uncs\ assigns greater weight to the SN~Ia \lh\ at lower redshifts
and is thus less sensitive to the \biasCor\ simulation at higher redshift.

% ======================================================
% ======================================================
% \bigskip
 \section{Results-II: More Realistic Tests }
 \label{sec:results-II}
% ======================================================
% ======================================================

Here we test the  \acro\  method on  realistic simulations that include 
CC \con, 
NN analysis,  
SN~Ia intrinsic scatter models with color variation  (G10 and C11 in \S\ref{sec:sim}), 
and Galactic reddening.
In addition to the flux \uncs, all light-curve fits include the \SALTII\ model \uncs,
which are based on the G10 model.
For the samples generated with the COH and C11 scatter models,
the incorrect G10-based model \uncs\ are still used in the light-curve fits.

For each of the three intrinsic scatter models,
 \NSAMPLE\ independent data samples are generated,
each with \NSIMDATA\ events passing selection requirements.
The averaged results are thus based on 150,000 events 
for each intrinsic scatter model.
A \biasCor\ sample with $\NSIMBIASCOR$ events is generated
for each intrinsic scatter model, and in the \acro\ fits we use
the \biasCor\ sample with the correct intrinsic scatter model.
Assuming the correct intrinsic scatter model is an optimistic assumption 
since we cannot clearly distinguish among  the G10 and C11 models,
but we leave these systematic tests to future analyses.

The data samples are generated with $\wdata = -1$ ($=\wref$),
which is very close to the assumption that $\wref$ has been set
to the first-iteration $\wfit$ value: a caveat is that we have not
included a statistical variation on $\wfit$.

The results are shown in Table~\ref{tb:fitResults_Ia+CC}.
The first three rows show results based on the  nominal CC rate ($\RCC=1$)
in the data and NN-training samples, and the resulting \con\ is $\sim 1$\%.
The next three rows show results with $\RCC=3$, resulting in 2.5\% contamination.
Compared to $\RCC=1$, the contamination with $\RCC=3$ is less than $3\times$ higher
because the NN training sacrifices a slightly  higher SNIa loss for a lower contamination.

\begin{table*}[t!]
\caption{ \acro\ Fit Results and $w$-Bias Averaged over $\NSAMPLE$ 
    Realistic DES-SN(Ia+CC) Samples\tablenotemark{a,b} }
\begin{center} \begin{tabular}{ | c | cc | cc | cc | c | cc | rc |  } \tableline  
 Intrinsic    & & True CC & & & & & & &  &  & \\
 Scatter Model    &  $\RCC$ & fraction\tablenotemark{c} & 
 $\alpha/\aSIM$  &  $\chisqred$ &
 $\beta/\bSIM$    &   $\chisqred$ &
 $\sigint$            &    
 $\SCC$             &  $\chisqred$ &
 $w$-bias  &  $\chisqred$ \\  [3pt]
   \hline\hline 
%
% first BCD section of BCD_results_Ia+CCx1.tex (Updated Nov 27 2016)
 %
COH & 1 & 0.009 & 1.005(4) & 0.9 & 0.992(2) & 1.1 & 0.109(3) & 1.02(4) & 1.7 & $ 0.015(4)$ & 1.2  \\  
G10  & 1 & 0.010 & 0.999(3) & 1.6 & 0.993(2) & 1.6 & 0.077(3) & 1.01(4) & 1.9 & $ 0.011(4)$ & 1.4  \\  
C11  & 1 & 0.010 & 1.015(3) & 1.2 & 0.990(2) & 1.1 & 0.100(8) & 1.08(4) & 1.3 & $-0.001(4)$ & 0.9  \\  

\hline
% first BCD section of BCD_results_Ia+CCx3.tex (updated Nov 27 2016)
%
COH  & 3 & 0.023 & 0.991(3) & 0.9 & 0.990(2) & 2.6 & 0.110(3) & 0.99(2) & 1.0 & $ 0.003(4)$ & 1.3  \\  
G10  & 3 & 0.025 & 0.999(3) & 0.4 & 0.991(2) & 1.9 & 0.076(3) & 1.01(2) & 0.9 & $ 0.004(4)$ & 0.8  \\  
C11  & 3 & 0.026 & 1.020(3) & 0.8 & 0.992(2) & 1.6 & 0.099(8) & 1.02(2) & 1.1 & $ 0.005(4)$ & 0.7  \\  
\tableline  \end{tabular} \end{center}   
\tablenotetext{1}{ Same as in Table~\ref{tb:fitResults1} }
\tablenotetext{2}{ Same as in Table~\ref{tb:fitResults1} }
\tablenotetext{3}{True $N_{\rm CC}/(N_{\rm CC}+N_{\rm Ia})$ after selection requirements.}
\label{tb:fitResults_Ia+CC} \end{table*}

The {\acro}-fitted $\alpha$ and $\beta$ are recovered to within 1\%, and the $w$-bias
is constrained at the level of 0.01.
The CC \prob\ scale factors, $\SCC$, are consistent with unity as expected.
Interestingly, the $w$-bias seems smaller with the larger CC rate,
albeit with low ($<2\sigma$) significance.  
The $w$-bias averaged over all six entries in Table~\ref{tb:fitResults_Ia+CC},
which includes \NSIMDATASUM\ simulated SNe, is \wBiasAvgCC.

To assess  the importance of the CC \lh\ in the \acro\ \lh\  (Eq.~\ref{eq:L}), 
we have  run the \acro\ fits without the CC term by forcing $\SCC=0$.
For the data sample generated with $\RCC=1$,  the fitted $\alpha$ and $\beta$ are 
biased by about 30\% and the $w$-bias is 0.1 to 0.15 depending on the intrinsic scatter model.
For the data sample with $\RCC=3$, the $w$-bias is around 0.2.
It is therefore essential to include an accurate CC term.

As an additional test on the impact of CC \con, we have  removed the
CC subset ($\RCC=0$), corresponding to a \specy\ confirmed (Ia-only) sample,
and fit with only the SN~Ia \lh.
For the COH and G10 scatter models, the Ia-only $w$-bias values are 
within $0.001$ of those obtained from the photometric sample with CC \con.
For the C11 scatter model, the $w$-bias values differ by 0.004.

A subtle issue is how the \acro\ fit further reduces the CC \con.
In \S\ref{sec:NN} (Table~\ref{tb:NN}) we showed that the NN requirement significantly 
reduces the CC contamination:
from \CCfracBoxCuts\ to \CCfracNN\ for $\RCC=1$, and
from \CCfracBoxCutsx\ to \CCfracNNx\ for $\RCC=3$.
However, the actual CC \con\ values after the \acro\ fit (Table~\ref{tb:fitResults_Ia+CC})
are about 50\% smaller: 0.01 and 0.025, respectively,
with a corresponding SN~Ia loss of 2\%.
After the NN requirement, the \acro\ fit further reduces the CC \con\ 
because the bias correction imposes an implicit selection requirement that is 
similar to the NN requirement, but  more strict.
Recall that the NN requirement rejects data events where more than 
half of the simulated neighbors are true CC SNe, and neighbor events
are counted within a 3D sphere defined by $\{z,x_1,c\}$.
The bias correction rejects events for which there are not enough SN~Ia 
events in a rectangular $\{z,x_1,c\}$-cell to determine the correction.
The \biasCor\ cell volume is $\sim 4\times$ smaller than the 3D sphere volume
determined from the NN training, 
and thus requiring a valid bias correction is more strict.

We further check the effectiveness of the \biasCor\ requirement in reducing 
CC \con\ by repeating the \acro\ analysis without the NN requirement, 
but still using the NN probabilities ($\PIa$ in Eq.~\ref{eq:L}).
With $\RCC=1$, the CC \con\ fractions are about 10\% higher than 
the combined NN+\biasCor\ requirement. 
With $\RCC=3$, the CC fractions are 20\% higher.
For both $\RCC$ values, the $w$-bias values are nearly identical
with and without the NN requirement.
Using only the NN requirement results in 50\% more CC \con, 
and therefore the \biasCor\ requirement is somewhat more effective 
than the NN requirement at reducing CC \con, 
but at a cost of 2\% additional loss of SN~Ia. 
This result should not be interpreted as a general statement
about the inferiority of the NN method,
because the NN method can also be used to obtain lower CC \con\
in exchange for reduced SN~Ia \eff.
For example, replacing  the ``purity$\times$\eff'' metric with
``purity$^p\times$\eff'' ($p>1$)  reduces the CC \con.

% ----------------------------------------------
\subsection{$\sigint$ from \acro\ Fit}
\label{subsec:fitsigma_int}
Before discussing the $\sigint$ results, it is useful to recall  two common methods
for characterizing intrinsic scatter in the HD.
First is the $\sigint$ method used here, which describes the additional scatter needed
to obtained a desired $\chisqHD$. From previous cosmology analyses on real data,
$\sigint \simeq \sigintVal$~mag \citep{Conley2011,JLA,BAHAMAS}; 
this value underestimates the true Hubble scatter because the \SALTII\ light-curve fit
includes model \uncs, derived from the training process, 
that increase the errors on $\{m_B,x_1,c\}$ and thereby reduces $\sigint$.
The other method is to report the rms of the Hubble residuals for a low-redshift 
sample with high $S/N$, and  results in typical values of 0.15~mag \citep{Jha2007}. 
The rms is not used in cosmology fits, but is a useful
metric for evaluating how well the SN~Ia brightness has been standardized.
While $\sigint$ does not include measurement \uncs, the residual rms does and 
is thus an overestimate of the Hubble scatter.

In the limit of infinite $S/N$ and zero \SALTII\ model \unc\ in the light-curve fits,
$\sigint$ and the Hubble residual rms are the same, and they are 
both equal to the $\sigcoh$ parameter for the COH model.
For the G10 and C11 intrinsic scatter models, $\sigint$ does not capture potential
dependences on color and stretch, but it still can be interpreted as the
average Hubble residual rms.   
This physical interpretation of $\sigint$ holds for realistic $S/N$ provided that
the light-curve flux \uncs\ are correct.
However, for realistic fitting with non-zero \SALTII\ model \unc, 
$\sigint$ is smaller than the Hubble scatter and therefore $\sigint$ 
does not have a physical interpretation.

In Table~\ref{tb:fitResults_Ia+CC}, $\sigint \sim\sigintVal$ for the COH model:
while consistent with previous analyses on real data,  it is 20\% smaller than
the input scatter value ($\sigcoh = \sigCOH$) used in the simulation. 
The presence of CC \con\ has no effect on $\sigint$, as verified by 
a \acro\ fit that excludes true CC and the CC \lh.
As described above, 
we expect that $\sigint < \sigcoh$ because of the \SALTII\ model \uncs.
To verify this effect, we refer back to our simplified simulation test (\S\ref{sec:results-I}) 
with the COH model and no \SALTII\ model \uncs: 
the {\acro}-fitted $\sigint$ agrees well with the input $\sigcoh$ value as shown in the
``$\sigint/\sigCOH$'' column in Table~\ref{tb:fitResults1}.

For the G10 scatter model, $\sigint \simeq \sigGTEN$~mag and is significantly 
below previous measurements;
this discrepancy suggests either a problem with the underlying model 
or with the implementation in the simulation. 
For the C11 model, $\sigint \simeq 0.1$~mag and is in good agreement
with previous measurements. 
To check the {\acro}-fitted values,
we have computed the true $\sigint$ value from low-redshift (high $S/N$) events by 
adjusting the rms of $(\mu_i - \mu_{\rm true})/\sigmui$ and using the true values of
$\alpha$ and $\beta$: the computed $\sigint$ agree well with
the {\acro}-fitted $\sigint$ for all intrinsic scatter models.

% ======================================================
% ======================================================
% \clearpage
% \vspace{0.5in}
 \section{Comparison with Other Bias-correction Methods}
 \label{sec:oldBiasCor}
% ======================================================
% ======================================================

Here we compare the \acro\ method with other HD-fitting methods.
First, in \S\ref{subsec:wrongFits} we show the impact from using
incorrect fitting methods: leaving out the \GN\ and/or the 
distance-bias corrections. 
Next, in \S\ref{subsec:previousBiasCor} we discuss previous
analyses which used a redshift-dependent distance-bias correction.

% -------------------------------------------------------------------
\subsection{Incorrect Fitting Methods}
\label{subsec:wrongFits}
% -------------------------------------------------------------------

Several recent SN~Ia analyses are based on the traditional $\chisqHD$ approach 
(Eq.~\ref{eq:chisqHD}) to HD fitting \citep{Conley2011,JLA,Rest2014,Scolnic2014b}.
While some approaches have included a redshift-dependent bias correction 
(e.g., \S\ref{subsec:previousBiasCor}),
the \GN\ term, $-2\ln(\sigma)$, has not been included because it results 
in large biases (e.g., see Appendix B of \citet{Conley2011}).  
The issue is explained in \cite{March2011}:
the \uncs\ on the color and stretch are comparable to the
natural width of the parent distributions, particularly at higher redshifts,
and therefore the Gaussian-error assumption in the $\chi^2$ \lh\ is not valid.
Their solution is to model the color and stretch distributions within the BHM framework.
Our \acro\ method uses bias corrections to model the correct mean in 
small 5D bins.

Within the \acro\ framework, 
both the \GN\ term and bias corrections are needed, 
and leaving out either results in a large bias, 
as shown in Table~\ref{tb:wrongFits}. 
The first two rows of Table~\ref{tb:wrongFits} show the effect of
including the \GN\ term without bias corrections: 
the fitted $\alpha$ and $\beta$ are $\sim 20$\% below their true values,
and the $w$-bias is greater than +0.1. 
The next two rows show the effect of including bias corrections 
without the \GN\ term: 
the fitted $\alpha$ and $\beta$ are $\sim 30$\% above their true values,
and the $w$-bias is almost $-0.1$.
The last two rows show the effect of leaving out the \GN\ and bias corrections:
the fitted $\alpha$ and $\beta$ are 5-10\% away from their true values
and the $w$-bias is $\sim 0.04$. 
It is interesting to note that leaving out either  the \GN\ or bias corrections
results in large fit biases with opposite signs; leaving out both corrections
results in some cancellation and  significantly smaller biases.
Using the  \acro\ method results in much smaller biases, 
as shown in Table~\ref{tb:fitResults1}.

\begin{table}[h!]
\caption{ HD Fit Results from Incorrect Fitting Methods\tablenotemark{a} }
\begin{center} \begin{tabular}{ | c |  c | c | c | r  |  } \tableline  
  Scatter     &  & &    & \\
  Model  & $\alpha/\aSIM$  &   $\beta/\bSIM$    &   $\sigint$     &     $w$-bias   \\  [3pt]
   \hline\hline 
%
% All incorrect results from  BCD_results_IaOnly_noCCx1.tex
%
%	 *****  chi2+ln(s) *****   
\multicolumn{5}{| c |  } {GN\tablenotemark{b}=yes ~~and~~ {\biasCor}=no} \\
  G10 & 0.787(3)  & 0.798(1)  & 0.103 & $ 0.118(4)$   \\  
  C11 & 0.787(3)  & 0.653(1)  & 0.120 & $ 0.126(4)$   \\  
\hline
%
%	 *****  chi2+biasCor ***** 
\multicolumn{5}{| c |  } {GN\tablenotemark{b}=no ~~and~~ {\biasCor}=yes} \\
 G10 & 1.281(4) &  1.269(2) &  0.066 & $-0.091(4)$   \\  
 C11 & 0.931(4) &  1.332(3) & 0.088 & $-0.085(4)$   \\  
 \hline
%
% chi2 only
\multicolumn{5}{| c |  } {GN\tablenotemark{b}=no ~~and~~ {\biasCor}=no} \\
 G10 & 1.069(4)  & 0.958(2)  & 0.091 & $ 0.037(4)$  \\  
 C11 & 1.083(4)  & 0.781(2)  & 0.110 & $ 0.038(4)$   \\  
\tableline  \end{tabular} \end{center}   
\tablenotetext{1} {Averaged over $\NSAMPLE$ simulated DES-SNIa samples without CC \con.}
\tablenotetext{2}{GN = \GN\ term  in HD fit.}
\label{tb:wrongFits} \end{table}

% -------------------------------------------------------------------
\subsection{ Redshift-dependent Bias Corrections }
\label{subsec:previousBiasCor}
% -------------------------------------------------------------------

\newcommand{\ab}{\{\alpha,\beta\}}

Some recent SN~Ia cosmology results  from large-survey teams
\citep{JLA,Scolnic2014b}
are based on the traditional $\chisqHD$ fitting approach with no \GN\ term,
but they  accounted for distance biases using simulations.
Recognizing that their HD fitting approach results in biased results,
their strategy was to measure the average $\mu$-bias ($\dmu$) as a function of redshift
by analyzing simulations with their biased HD fitting procedure
that has no \GN\ term and no bias corrections.
Each distance modulus in the data is then corrected as a function of redshift,
$\mu \to \mu -\dmu(z)$, which we call a 1D \biasCor\ to distinguish from the
5D \biasCor\ in the \acro\ method. The final cosmology fit is applied to the
1D-corrected distances.

With an unrealistic assumption of using
the correct $\alpha$ and $\beta$ in the \biasCor\ simulation,  
the 1D \biasCor\ works as well as the 5D \acro\ method,
but with 9\% larger \unc\ on cosmological parameters.
An important caveat is that previous analyses did not have a robust method 
for determining $\alpha$ and $\beta$, which are needed as inputs to the
\biasCor\ simulation.
In \citet{JLA}, $\alpha$ and $\beta$ were determined from the
traditional $\chisqHD$ fit, and thus may have  biases as
illustrated for DES in the last two rows of Table~\ref{tb:wrongFits}.
\citet{Scolnic2014b,Scolnic2014a} used a similar approach and analyzed additional simulations
to correct for the large $\beta$-bias in the C11 model.

To illustrate the impact of incorrect $\alpha$ and $\beta$,
we generated a \biasCor\ simulation with 
$\alpha=0.15$ and $\beta=3.0$, which corresponds roughly
to the biased values using the traditional $\chisqHD$ approach 
(last 2 rows in  Table~\ref{tb:wrongFits}). 
We use this \biasCor\ simulation to apply a 1D correction
to the simulated data samples from \S\ref{sec:results-I}
that were generated with the true parameter values ($\alpha=\agen$, $\beta=\bgen$):
the resulting $w$-bias is $0.016 \pm 0.004$, significantly larger than the
5D \acro\ result in the first row of Table~\ref{tb:fitResults1}.

The fundamental issue is not so much the 1D-vs-5D correction,
but rather that previous analyses bias-corrected the distance  ($\mu\to\mu-\dmu$) 
instead of correcting each fitted parameter ($m_B,x_1,c$) 
shown in Eq.~\ref{eq:mucor} for the \acro\ method.
The $\dmu$ correction requires $\alpha$ and $\beta$ to be predetermined,
while the \acro\ method explicitly updates $\dmu$  at each minimization step
as $\alpha$ and $\beta$ are varied.

It would be interesting to apply the \acro\ method to existing data samples
in order to measure $\alpha$ and $\beta$,  
and to check for potential errors on the 1D distance-bias corrections and
cosmological parameters. 
However, evaluating the $\alpha,\beta$ bias in previous analyses is beyond the
scope of this work.

% ======================================================
% ======================================================
% \clearpage 
 \section{Discussion and Conclusion}
 \label{sec:fin}
% ======================================================
% ======================================================

The \acro\ fit has been implemented within the existing \Smu\ program;
it is publicly available in \SNANA, and ready to use on real data samples for 
which reliable simulated samples are available. 
This new program has been rigorously tested on large simulated 
data samples ($\sim 150,000$ events) with different models of
intrinsic scatter and different CC rates.
The CPU time for each \acro\ fit  is less than 10~minutes for 10,000 events,
and this time scales with the number of events.
A \biasCor\ simulation sample of \NSIMBIASCOR\  events ($z<1.2$) 
takes about 50 CPU hours to create,\footnote{
The \biasCor\ CPU breakdown is 10~hr for SN~Ia generation, 
and almost 40~hr for light-curve fitting. }  % end footnote
and a few hours are needed to generate a Ia+CC sample and perform
the NN training.
Below we describe some unresolved issues and areas for  improvement.

The first unresolved issue is how to determine sub-samples
for which the \biasCor\ simulation is run. 
The deep and shallow fields in \DESSN, with a 1~mag difference in 
search depth, is a rather obvious choice for separating the samples.
However, for a survey with a fixed exposure time there are
weather variations which change the search depth, 
and edge effects that truncate light curves; 
it is thus not clear if or how the \biasCor\ simulation should be split
into multiple samples. 
Our current method for splitting samples is essentially trial-and-error. 
In this analysis, for example, we started with shallow and deep field samples  
combined into a single \biasCor\ simulation, and found $w$-biases 
in the range of 0.01 to 0.02; generating separate \biasCor\ simulations
for shallow and deep fields reduced the $w$-bias to below 0.01.

Most previous cosmology \lhs\ are based on summing a contribution
from each event where the size of covariance matrix ($\COV$)
grows as $\Ndata^2$. As data samples increase into the tens of thousands,
there would likely be  computational challenges in both memory and fitting speed.
The \acro\ approach results in a $z$-binned cosmology \lh\
(Eq.~\ref{eq:chisqDelta}) that sums over  $\Nz$ redshift bins. 
The corresponding covariance matrix ($\COV$)
has a  manageable  and fixed size of $\Nz \times \Nz$, 
and should not introduce computational problems for large data sets.
A $z$-binned cosmology fit, including systematic \uncs, 
has recently been demonstrated in Appendix~E of \citet{JLA}.
However, further testing is needed to validate a
$z$-binned $\COV$ within the \acro\ framework.

Perhaps the most worrisome long-term issue is whether intrinsic scatter
can be adequately described by the $\sigint$ term, or even a more complex 
$3\times 3$ intrinsic scatter matrix ($\Cint$) in the space of $\{m_B,x_1,c\}$.
The \acro\ and BHM methods can be enhanced to fit for additional $\Cint$
parameters, but the sensitivity is unclear, nor do we know what
range of wavelength-dependent and time-dependent variations can be
reasonably captured by $\Cint$.  
In our simulation tests, the G10 and C11 scatter models are implemented 
with spectral variations, with no analytical assumption about $\Cint$, 
and the $\sigint$ description in the \acro\ fit works surprisingly well,
resulting in a $w$-bias below 0.01. Continued tests of this nature
will be essential to validate the simplified intrinsic scatter model
in the Hubble fitting \lh.  
It is worth noting here that the ABC method \citep{ABC2013,ABC2016}
has no \lh\ and can therefore accommodate an arbitrarily complex model of intrinsic scatter.

Another issue is that \uncs\ on the assumed cosmology in the \biasCor\ simulation
introduce an additional $w$-\unc\ of $\sigwAdd$ for the \DESSN\ sample
(\S\ref{sec:results-I} and Fig.~\ref{fig:wfit_test}).
This finding raises the prospect that the highest redshift events could add 
a \biasCor\ \unc\ that  exceeds the reduced statistical \unc.
This bias has been ignored in all previous analyses
that use a \biasCor\ simulation,
although our analysis suggests that the impact is small compared to the total \unc.

An obvious solution is to expand the dimensionality of the 
\biasCor\ grid to include cosmology parameters, and to include 
these cosmology parameters in the \acro\ fit. The downside of this solution
is that external HD fitting codes cannot be used, and as explained above,
a large ($\Ndata^2$) covariance matrix would be needed for systematic \uncs.
To preserve the feature of creating a binned HD for
external fitting codes, another solution is to generate
\biasCor\ samples for a grid
of cosmological parameters ($\COSVEC$) 
such as $\COSVEC = \{w,\OM\}$ or $\COSVEC = \{w_0,w_a,\OM\}$,
where $w(a) = w_0 + w_a(1-a)$ and ``$a$'' is the scale factor for
the expanding universe.  Generating a \biasCor\ grid with two values per
cosmology parameter requires $2^2$ and $2^3$ \biasCor\ simulations
for the two models, respectively, which only requires a few hundred CPU hours.
Arbitrary cosmology fitting programs could  interpolate the 
{\acro}-fitted $\DMUz$ as a function of $\COSVEC$,
with the limitation of using only the cosmological model defined by $\COSVEC$.
While a truly model-independent output from \acro\ is desired, it is not clear how to 
achieve this.

Using a BHM  framework, we mention another possible 
solution to the cosmology-dependent bias correction.
In the UNITY method of \citet{UNITY} they avoid simulations and instead
characterize the \eff\ as an ad hoc function of \SALTII\ parameters.
The \eff\ dependence on $m_B$ should include the cosmology dependence.
It is not clear, however, what \uncs\ are introduced by the \eff\ function.

A number of enhancements to \acro\  are feasible with minor code changes.
Additional fitted parameters in the CC component would add flexibility,
such as including a redshift-dependent $\sigint$ parametrization, and 
allowing the Gaussian mean and width in each redshift bin (Fig.~\ref{fig:CCmap}) 
to be adjusted by a polynomial function of redshift.
Fitted polynomial coefficients different from zero would indicate
a discrepancy between the data and CC simulation.
Instead of using the simulated CC map, redshift-dependent parametrizations in 
\citet{Jones2016} can also be implemented.
The SN~Ia \lh\ could be enhanced with additional parameters to describe 
host-galaxy correlations, redshift-dependent SN properties, 
and quadratic stretch and color terms in the Tripp equation.
Last, the interpolated bias corrections are linear between the nearest 
5D grid nodes where the biases are defined.
Spline interpolation may work better, as long as the CPU time and memory
are not significantly increased.

The population parameters describing the asymmetric color and stretch 
distributions have been determined in a separate analysis (SK16).
However, it may be possible to include these parameters in the \acro\ fit
by multiplying the \lh\ by a simulated \prob\ characterizing the number
of events in stretch and color bins. The downside of this enhancement
is that the \SALTII\ parameter biases cannot be determined before the fit, 
and thus could result in a significant increase in CPU time.
While a simultaneous fit of all parameters is an attractive goal,
it may be more computationally efficient to iteratively evaluate the population 
parameters separately from the \acro\ fit.

In summary, we have presented a new SN~Ia  cosmology fitting method, 
\acro,  which uses a large simulation to account for biases from 
sample selection, 
light-curve fitting,
and CC contamination. 
Analyzing nearly a million simulated supernova, we find that  
the \acro\ method introduces a $w$-bias below 0.01.
A proper cosmology analysis, however, should characterize and account for
\uncs\ on the \biasCor\ and CC simulations.

% ======================================================
% ======================================================
 \section{ Acknowledgements }
 \label{sec:Ack}
 
This work was supported in part by the Kavli Institute for Cosmological Physics at the University of Chicago 
through grant NSF PHY-1125897 and an endowment from the Kavli Foundation and its founder Fred Kavli.
R.K. is supported by DOE grant DE-AC02-76CH03000.
D.S. is supported by NASA through Hubble Fellowship grant HST-HF2-51383.001 awarded by the
Space Telescope Science Institute, which is operated by the Association of
Universities for Research in Astronomy, Inc., for NASA, under contract NAS 5-26555.

We gratefully acknowledge support from NASA grant 14-WPS14-0048.  
This manuscript is based upon work supported by the National
Aeronautics and Space Administration under Contract No.\ NNG16PJ34C
issued through the {\it WFIRST} Science Investigation Teams Program.

While thanking the anonymous referee is good common practice, 
we are exceptionally grateful to our referee and feel fortunate 
to have received valuable contributions.

% ==============================================================
% ==============================================================
% BIBLIOGRAPHY
% \clearpage

\bibliographystyle{apj}
\bibliography{BBC_ms}  

% ==============================================================
% ==============================================================

% ####################################
  \end{document}